\documentclass[12pt]{article}
\usepackage{amsmath}
\usepackage{amssymb}
\usepackage{graphicx}
\usepackage{color}
\usepackage{psfrag}
\usepackage{verbatim}
\usepackage[tight,sf]{subfigure}
\usepackage{hyperref}

\newcommand{\VECb}{{\boldsymbol{b}}}
\newcommand{\VECe}{{\boldsymbol{e}}}
\newcommand{\VECu}{{\boldsymbol{u}}}
\newcommand{\VECt}{{\boldsymbol{t}}}
\newcommand{\VECn}{{\boldsymbol{n}}}
\newcommand{\VECi}{{\boldsymbol{i}}}
\newcommand{\VECj}{{\boldsymbol{j}}}
\newcommand{\VECk}{{\boldsymbol{k}}}

\newcommand{\VECJ}{{\boldsymbol{J}}}
\newcommand{\VECX}{{\boldsymbol{X}}}

\newcommand{\CALK}{{\mathcal{K}}}
\newcommand{\ellipticK}[1]{\CALK\left[ #1\right]}
\newcommand{\Jacobisnadjustroundb}[2]{\operatorname{sn}\left(  #1,#2 \right)}
\newcommand{\Jacobisnadjustsquareb}[2]{\operatorname{sn}\left[  #1,#2 \right]}
\newcommand{\Jacobiamadjust}[2]{\operatorname{am}\left(  #1,#2 \right)}

\newcommand{\se}{\varphi_{\text{e}}}
\newcommand{\st}{s_{\text{e}}}

\begin{document}

\sf
\begin{center}
   \vskip 2em
     {\Large \sf
     CONICAL INSTABILITIES ON PAPER}
  \vskip 3em
  {\large \sf
Jemal Guven$^1$, Martin Michael M\"uller$^2$ and Pablo V\'azquez-Montejo$^1$
 \\[2em]}
\em{$^1$ Instituto de Ciencias Nucleares,
 Universidad Nacional Aut\'onoma de M\'exico\\
 Apdo. Postal 70-543, 04510 M\'exico, D.F., MEXICO\\[1em]
$^2$ Equipe BioPhysStat, ICPMB-FR CNRS 2843, Universit\'{e} Paul
Verlaine-Metz\\
1, boulevard Arago, 57070 Metz, FRANCE
}
\end{center}
 \vskip 1em

\begin{center}
\today
\end{center}

\begin{abstract}
\sf
The stability of the fundamental defects of an unstretchable flat sheet is examined. 
This involves expanding the bending energy to second order in deformations about the defect. The modes of deformation occur as 
eigenstates of a fourth-order linear differential operator. Unstretchability places a global linear constraint on these modes.
Conical defects with a surplus angle exhibit an infinite number of states. If this angle is below a critical value, these states possess an 
$n$-fold symmetry labeled by an integer, $n\ge 2$.  A nonlinear stability analysis shows that the 2-fold ground state is stable, whereas excited states 
possess $2(n-2)$ unstable modes which come in even and odd pairs. 
\end{abstract}


\section{\sf Introduction}

\vskip1pc\noindent
In general, when an elastic sheet bends it will also stretch. While the energy associated with bending is
purely geometrical--depending on the surface curvature--the energy associated with stretching will
depend on additional material degrees of freedom. 
However, if the sheet is very thin compared to a typical radius of curvature,
stretching will typically become far more costly than bending; thus, if stretching does
occur, it will tend to be confined within a set of sharp peaks connected by a network of folds in
the sheet \cite{Fuchs,BenPom,cerdamaha}. This observation has attracted many physicists in the last decade, who have started to look at the properties of these defects in detail (a nice review is provided in reference \cite{Witten}). 
The limiting behavior of thin sheets has one striking feature: it involves only the
geometrical degrees of freedom of the sheet, in particular the bending energy depends only on its
shape;
unstretchability constrains what shapes are accessible.  This simple picture turns out
to provide a description of the behavior of a range of real materials on widely differing scales: it
is equally valid for a sheet of graphene on microscopic scales as it is for a sheet of paper on
macroscopic ones \cite{Witten,RMPGraphene}. 
Corresponding materials of soft matter have been studied only recently, mainly inspired by biology. 
It was shown for example, that growing tissues or thermo-responsive 
gels can be described within the same limit if they are sufficiently thin \cite{Sharon,DervauxBenAmar,econe}. 
A general treatment of their stability is, however, still lacking. The goal of this paper is to analyze the stability of the basic building block of such materials, the conical defect. 

\vskip1pc\noindent
Approaching the stability of the equilibrium states of an unstretchable sheet is complicated by
the local nature of the constraints. These constraints distinguish their treatment from 
that of a fluid membrane which can also be modeled geometrically \cite{CanHel}.  In fluid membranes
the penalty associated with shear is very small so that it behaves like a fluid along tangential
directions.  If there are constraints in the latter they tend to be global in nature: area, enclosed
volume or--in a bilayer--area difference between inner and outer layer may be constrained
\cite{Seifert}. In an unstretchable sheet, however, the constraints are local.  Unstretchability
translates into a constraint on the metric tensor induced on the surface \cite{Paperfold}. While the
two limits lend themselves to a geometrical description the physical behavior they display is qualitatively
very different. This is reflected in how one goes about examining this behavior.

\vskip1pc\noindent
The metric constraint lends rigidity to the surface with respect to its behavior as a fluid
membrane. Whereas a shape that is stable as a fluid membrane will also be stable if it is
unstretchable, the converse is not true:  unstable modes of a fluid membrane will tend to become
inaccessible when the constraint on the surface is accommodated.  As a case in point, there are
conical defects that are generic stable features of unstretchable flat sheets.  If the apex of the
cone is part of the surface it is not possible to smooth it out by surface deformations that
preserve the conical metric. In a fluid membrane, however,  the energy will be lowered by smoothing
out the singularity at the apex. Thus, conical defects will never appear as stable features of a fluid
membrane. In this paper the discrepancy will be quantified.

\vskip1pc\noindent
The stability of simple fluid membrane equilibrium states--spheres and cylinders--was addressed by
Helfrich and Ou-Yang in the late 80s \cite{HOY}.  This involves the evaluation of the second
variation of the bending energy about these surface states. This was a non-trivial calculation and,
not surprisingly, their approach was adapted specifically to these states. Later a framework was
developed which does not depend explicitly on the equilibrium state \cite{CGS}.

\vskip1pc\noindent
To accommodate the unstretchability of the sheet, we will adopt the method of Lagrange multipliers
in the calculus of variations to enforce the corresponding metric constraint.  Like the construction
for fluid membranes developed in \cite{CGS} the framework we will develop will not depend explicitly
on the particular equilibrium shape. Our focus will be on flat sheets, which can be modeled as surfaces 
with vanishing Gaussian curvature, and we will derive an explicit expression for the second variation of the bending energy
subject to this constraint. The unstretchability manifests itself in the important respect that the
second variation, unlike its fluid membrane counterpart, depends now on both the normal and
tangential deformations of the surface.  This is, as we will see, because the constraint spoils the
identification of tangential deformations with reparametrizations.  That an exact framework exists
at all is somewhat surprising; that the calculation is largely tractable analytically is even more
so.

\vskip1pc\noindent
This geometrical framework will be used to examine the stability of the equilibrium states of a
circular flat sheet with a conical defect at its center. Such surfaces are in one-to-one
correspondence with closed curves of fixed length on the unit sphere.  We will focus, in particular,
on a sheet with a surplus angle at this point \cite{DervauxBenAmar,econe} (see also \cite{Warner} for a related example involving nematic solids).
Unlike their counterparts with an angle deficit such cones possess an infinite number of equilibrium states in the absence of external forces.  If the surplus angle is sufficiently small, so that self-contacts do not occur,  these shapes are characterized by an $n$-fold symmetry beginning with a $2$-fold. To assess their stability we obtain an explicit expression for the second variation of the energy about each of these states. This will involve a self-adjoint fourth-order linear differential operator, ${\cal L}$, which depends explicitly on the curvature of the conical state. The accessible modes of deformation are described by the eigenmodes of this operator consistent with a global linear constraint associated with the unstretchability.
A negative eigenvalue signals an instability. If the surplus angle is small this constraint is tractable analytically and an exact Fourier analysis of stability is possible.

\vskip1pc\noindent  The rotational invariance of the energy implies the existence of three zero modes of deformation for
each equilibrium state of the conical defect. An exact expression for these modes will be provided. Two of them correspond to rotations about axes orthogonal to the axis of symmetry. For small surplus angles they possess a pair of nodes.  The third zero mode is odd and corresponds to a rotation about this axis; it has $2n$ nodes in this limit.  All eigenmodes with an intermediate number of nodes will be unstable.

\vskip1pc\noindent  In general, an exact treatment of stability
eludes us.  So we develop an approximation for larger surplus angle by introducing
a Fourier expansion of the modes; using the Gram-Schmidt process we construct an orthonormal basis satisfying the global constraint. The matrix elements of the linear operator are evaluated.
While the ground state is stable--like its ice-cream counterpart--as one would have predicted, all excited states are unstable. The detailed nature of this instability is, however, rather interesting. The $n$-fold beginning with $n=3$ will possess $2(n-2)$ unstable modes.  These modes occur in pairs with even and odd parity with respect to the $n$-fold.  For low surplus angles, each mode consists of a single frequency. As the surplus angle increases higher frequencies enter and the number of nodes of each mode increases.  It is nevertheless possible to order modes in a manner analogous to the small surplus angle limit. The spectrum of the operator ${\cal L}$ will be discussed and the dominant mode of instability will be identified for each excited state.


\section{\sf Second order variations with a metric constraint} \label{2ndvar}

We will model the unstretchable sheet in terms of a surface which is described parametrically by three functions $\VECX(u^1,u^2)$,
$\VECX= (X^1,X^2,X^3)$  providing its position in three-dimensional
space. The bending energy associated with the folded state of the surface
is quadratic in curvature,
\begin{equation}
H_B[\VECX] = \frac{1}{2}\int dA\,\, K ^2 \, ,
\end{equation}
where $dA$ is the area element induced on the surface and $K =
C_1 + C_2$, where $C_1$ and $C_2$ are the two principal curvatures.
$H_B[\VECX]$ thus involves only the geometry of the surface.
Once the geometry is specified, the bending energy is also.
We will work in units in which the rigidity modulus is unity.

\vskip1pc\noindent
The induced  metric tensor on the surface is given by $g_{ab}= \VECe_a\cdot \VECe_b$, where $\VECe_a
= \partial_a \VECX$, $a=1,2$,  are the two tangent vectors of the surface adapted to the
parametrization by $u^1$ and $u^2$. This parametrization will be fixed once and for all.
Geometrically, the unstretchability of the sheet is then the
statement that the surface is isometric, $i.e.$, the only admissible deformations are those keeping $g_{ab}$ fixed.
A natural way to keep track of this constraint in the calculus of variations
is to replace $H_B[\VECX]$ by the functional \cite{Paperfold}
\begin{equation}
H[\VECX, T^{ab}]= H_B[\VECX] - \frac{1}{2}\,\int dA\, T^{ab}
\, (g_{ab}- g_{ab}^{(0)})\,.
\end{equation}
The symmetric tensor $T^{ab}$ is the  set of
Lagrange multipliers  associated with the constraint that
$g_{ab}$ coincides with some fixed metric $g_{ab}^{(0)}$.  In particular,
we will be interested in a surface that is a circular flat disc with a conical singularity, described by the line element
\begin{equation}
dl^2 = dr^2 + r^2 ds^2\,,
\end{equation}
where $r$ is the distance to the apex of this cone and $s$ ranges between 0 and some value $2\pi + \se$.
However, our framework--at least in its initial stage--will not require us to specify
$g_{ab}^{(0)}$ explicitly.

\vskip1pc\noindent
Recall that a surface is completely determined once the metric tensor $g_{ab}$ and the extrinsic
curvature tensor  $K_{ab}= \VECe_a\cdot \partial_b \VECn$ are specified \cite{Spivak,Montiel}. The
symmetric second rank tensor $K_{ab}$  is a measure of how fast the normal vector $\VECn$ rotates
into one  direction as it is moved along another. Unlike $g_{ab}$, it depends explicitly on this
vector. The principal curvatures, $C_1$ and $C_2$, are the maximum and minimum values of the shape operator $K_{ab}$. They are
assumed along two tangents--the principal directions--which are always perpendicular. Fixing the
metric places a constraint on deformations of the surface that is more severe than may, at first,
appear. A familiar example is provided by the pingpong ball in which curvature and spherical
topology together conspire to produce a geometry that is rigid under isometric deformations. 

\vskip1pc\noindent In order to say something about stability we first identify the Euler-Lagrange equations that describe the equilibrium states subject to the constraint.
Consider a deformation $\VECX\to \VECX+ \delta \VECX$, $T^{ab}\to T^{ab} + \delta T^{ab}$.  As described in reference
\cite{Paperfold},  where the variational framework developed in references
\cite{stress} and \cite{auxil} was adapted to accommodate the metric constraint, the corresponding change in $H$ at first order is given, modulo a boundary term, by
\begin{equation}
\label{eq:delH}
\delta H = \int dA\, ({\cal E}_\perp \VECn + {\cal E}_\|^b {\VECe}_b) \cdot \delta \VECX - {1\over 2}\, \delta T^{ab} (g_{ab}-
g_{ab}^{(0)})\,,
\end{equation}
where
\begin{eqnarray}
{\cal E}_\perp  &=&  -\, \nabla^2 K + \frac{1}{2} K (4 K_G -
K^2 )
 - K_{ab} T^{ab}\,,
\label{eq:EperpX}\\
{\cal E}^a_\| &=& \nabla_b T^{ab}\,.
\end{eqnarray}
Here $K_G=C_1 C_2$ is the Gaussian curvature.
The constraint adds a term linear in curvature to the normal Euler-Lagrange derivative ${\cal E}_\perp$. Unlike a fluid membrane, the tangential counterpart is not identically zero.  The equilibrium of the sheet is then described by the Euler-Lagrange equations:
\begin{equation}
{\cal E}_\perp = 0\,,\quad {\cal E}^a_\|=0\,,\quad g_{ab}=
g_{ab}^{(0)}\,.\label{eq:EL3}
\end{equation}
In equilibrium, the tangential stress $T^{ab}$ associated with the metric constraint  is conserved. This equation involves only the intrinsic geometry. The extrinsic geometry enters through the boundary conditions on the free surface of the sheet.
In general, these conditions will involve the introduction of a boundary layer
in which stretching is admitted \cite{Witten,Sharon}.  Fortunately, in the case we will treat here, the energy itself will be independent of the precise nature of these conditions. The only  component of $T^{ab}$ which enters the stability analysis is the trace $T^{ab} K_{ab}$ coupling to curvature in ${\cal E}_\perp$.
In \cite{Paperfold} it was shown that this component is insensitive to boundary conditions.

\vskip1pc\noindent We now need to extend this framework
to address the stability of equilibrium shapes. To do
this it is necessary to expand $H$ to second order about solutions of the
Euler-Lagrange equations. Note that when the Euler-Lagrange equations are satisfied,
\begin{equation}
\label{eq:del2H0}
\delta^2  H = \int dA\, \delta {\cal E}_\perp\, (\VECn \cdot
\delta \VECX) + \delta {\cal E}^a_\|\, (\VECe_a \cdot \delta
\VECX)\,.
\end{equation}
Here
\begin{eqnarray}
\delta {\cal E}_\perp &=& \delta_\VECX {\cal E}_\perp - K_{ab}
\delta T^{ab}  \,; \label{eq:delEperp}\\
\delta {\cal E}^a _\| &=& \nabla_b \delta T^{ab} \,,
\end{eqnarray}
where $ \delta_\VECX {\cal E}_\perp $ represents the first order change in 
${\cal E}_\perp$ given by Eq.(\ref{eq:EperpX}) under the isometry $\delta \VECX$,
maintaining  $T^{ab}$ fixed. 
Let us examine $\delta {\cal E}^a _\|$. Note first that terms involving the deformed
Christoffel connection $\Gamma^a_{bc}$ vanish whenever $\delta g_{ab}=0$.
A more general statement is possible: whenever $\delta
g_{ab}=0$, the variation commutes with covariant differentiation.
We now integrate by parts to express the second term appearing in Eq.(\ref{eq:del2H0})
in the form
\begin{equation}
\label{eq:nabT}
\nabla_b \delta T^{ab}\, (\VECe_a \cdot \delta \VECX) = \delta
T^{ab} K_{ab}\,(\VECn\cdot \delta \VECX) - \delta
T^{ab}\,(\VECe_a\cdot \nabla_b \delta \VECX)\,,
\end{equation}
modulo a divergence.
The first term cancels an identical term  appearing in $\delta
{\cal E}_\perp$ (see Eq.(\ref{eq:delEperp})); the second term vanishes
because
\begin{equation}
\label{delg0}
\delta g_{ab}= \VECe_a\cdot \nabla_b \delta \VECX+
\VECe_b \cdot \nabla_a \delta \VECX =0
\,.
\end{equation}
What is left is the surprisingly simple expression 
\begin{equation}
\label{eq:del2HX}
\delta^2  H = \int dA\, \delta_\VECX {\cal E}_\perp \, (\VECn
\cdot \delta \VECX) \,.
\end{equation}
Superficially, it might
appear that the second order deformation is identical to
the one that would have been obtained (see reference \cite{CGS})
had $T^{ab}$ been treated as a parameter. All that one seems to have achieved is the satisfaction of having dotted all of the i's, confirming that what one might have
guessed is also correct. The presence of the constraint, however,  spoils the identification of
tangential deformations with reparametrizations of the surface: in particular, a tangential
deformation of ${\cal E}_\perp$ does not vanish. While ${\cal E}_\perp$ is a
surface scalar, it involves the tangential stress $T^{ab}$ which does not depend explicitly on $\VECX$.
It is thus not completely determined by the local surface geometry.\footnote {\sf A small tangential
deformation of any geometrical tensor whose spatial dependence is determined locally by $\VECX$ can be identified
with a reparametrization of the surface (a Lie derivative). It thus vanishes when its argument vanishes.}
As a consequence  $\delta_\VECX {\cal E}_\perp$ is not the same as $\delta_\perp {\cal E}_\perp$ in an equilibrium state, and it is no longer legitimate to replace one by the other as it was in the case of a
fluid membrane (see, for example, \cite{CGS}).

\vskip1pc\noindent  While this does not appear to bode well, the evaluation of
$\delta_\VECX {\cal E}_\perp$ turns out to be simpler than that of
$\delta_\perp {\cal E}_\perp$. This is because it is possible to exploit
the fact that isometric variations commute with covariant differentiation.
At a later stage in the calculation, it will be necessary to implement the
metric constraint explicitly to eliminate tangential deformations in favor of normal ones.

\vskip1pc\noindent
To obtain
Eq.(\ref{eq:del2HX}) we discarded a boundary term in Eq.(\ref{eq:nabT})
involving $\delta T^{ab}$. Had we kept track of it,
we would have seen that it cancels a counterpart originating in
the boundary term that tags along on the right hand side of Eq.(\ref{eq:delH}) when derivatives of $\delta \VECX$ are collected in a divergence.
There remains to determine $\delta_\VECX {\cal E}_\perp$
appearing in Eq.(\ref{eq:del2HX}).

\vskip1pc\noindent It is useful to think of the normal Euler-Lagrange derivative,
${\cal E}_\perp $, defined by Eq.(\ref{eq:EperpX}) as a sum of two terms.  The bending part
depends on the extrinsic geometry only through the mean curvature. Using the fact that both
 the Laplacian and the Gaussian curvature
are invariants under isometry, it is simple to see that
\begin{equation}
\delta_\VECX {\cal E}_\perp = - \nabla^2 \delta K + 2 K_G \, \delta K -
{3\over 2} K^2 \delta K - T^{ab} \delta K_{ab}\,.
\end{equation}
The bending contribution only requires the evaluation of $\delta K$.
The tensor $\delta K_{ab}$ does, however, enter the
source term associated with the constraint.
In appendix~\ref{app:1stvariationKKab}, $\delta K_{ab}$ and $\delta K$ are given in terms of the deformation $\delta\VECX$. In the following we will specialize the obtained expressions to study the stability of conical defects on a flat unstretchable sheet.


\section{\sf Conical Deformations}

\subsection{\sf Cones as curves on spheres}

\begin{figure}
\begin{center}
  \includegraphics[bb=100 550 400 675,width=0.6\textwidth]{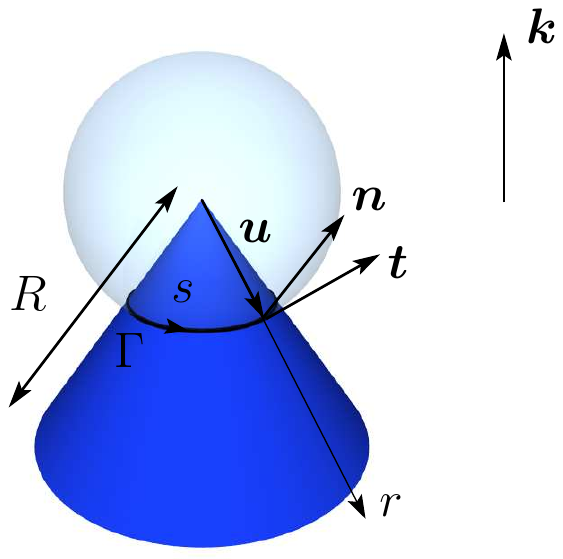}
\end{center}
\caption{The geometry of the conical defect.}\label{fig:econe}
\end{figure}

We will use the representation of the conical defect as a closed curve  $\Gamma: s\mapsto \VECu(s)$
on a unit sphere (see Fig.~\ref{fig:econe}). If $r$ is the distance to the apex, then
a parametric description of the cone is provided by the identification $\VECX(r,s) = r \VECu(s)$.
It is convenient to parametrize the curve by arc-length. The total length of the curve $\st=2\pi + \se$ will be invariant under isometry. Indeed, the intrinsic geometry of the cone
is characterized completely by $\st$. For technical simplicity,  let us suppose that
this cone has a fixed finite radius $R$. If $\st=2\pi$, this cone will be isometric to  a planar disc.  If $\st< 2\pi$, it will exhibit an angle deficit at its apex. The unique equilibrium configuration in the absence of external forces is then an axially-symmetric ice cream cone (see again Fig.~\ref{fig:econe}).
Our focus will be on cones with a surplus angle $\st>2\pi$. Unlike their counterparts with an angle deficit, as described in \cite{econe}, there will be an infinite number of non-trivial equilibrium states.  In particular, if the surplus angle is sufficiently small so that self-contacts do not occur, these states will be symmetrical with an $n$-fold symmetry labelled by an integer $n\ge 2$. The axis of symmetry will be aligned with the basis vector $\VECk$ of the Euclidean coordinate system $(\VECi,\VECj,\VECk)$. Once self-contact occurs the symmetry of the cone is broken and skewed geometries emerge \cite{Stoop2010}.

\vskip1pc\noindent
If $\VECt$ is the tangent vector to the curve $\VECu(s)$, and $\VECn=\VECu\times \VECt$, then
$\{\VECu,\VECt,\VECn\} $ forms an orthonormal frame satisfying
\begin{equation}\label{frameprime}
\VECu'=\VECt\,,\quad
\VECt'= -\kappa \VECn - \VECu\,,\quad \VECn'= \kappa \VECt\,,
\end{equation}
where $\kappa$ is the geodesic curvature of the curve on the sphere (which, confusingly,  turns out to be its normal curvature if it is viewed as
a curve on the cone).

\vskip1pc\noindent The extrinsic curvature tensor is given by
\begin{equation} K_{ab}=  r \left( \begin{array}{cc}
  0 & 0\\
  0 & \kappa
  \end{array}\right)
\,.\label{eq:Kab}
\end{equation}
The flat direction is along $\VECu$.  The Gaussian curvature
$K_G$ vanishes.
The trace $K= \kappa/r$
diverges at the apex of the cone. This translates into a logarithmic
divergence in the bending energy at this point.

\vskip1pc\noindent The normal projection of the Euler-Lagrange
derivative is given by
\begin{equation}
{\cal E}_\perp =
- \frac{1}{r^3} \left[\kappa'' + {1\over 2} \kappa^3 +
(1-C_\|) \kappa \right]
\,, \label{eq:Eperpcone}
\end{equation}
where
$C_\|=- r^4 T^{ss}$ is a constant related to the stress along the tangent direction $\VECt$ \cite{Paperfold}.
The vanishing of ${\cal E}_\perp$ determines the equilibrium shapes of the cone. Its first integral
\begin{equation}
 J^2 - C_{\|}^{2}
 = \kappa'^2 + \frac{1}{4}\kappa^4 + (1 - C_\|) \kappa^2
  \label{eq:firstintegral}
\end{equation}
can be solved in terms of elliptic functions where $\VECJ = J\VECk$ is a constant vector aligned with $\VECk$
which is associated with the rotational invariance of the conical bending energy \cite{Paperfold}. It is given by
\begin{equation}
  \VECJ = \left(\frac{\kappa^2}{2}-C_\|\right)\VECn + \kappa' \VECt + \kappa \VECu
  \; .
  \label{eq:VECJ}
\end{equation}
Solving the first integral~(\ref{eq:firstintegral}) for a cone with surplus angle $\se$ and $n$ folds one obtains \cite{econe}
\begin{equation}
  \kappa (s) = \frac{4\sqrt{-k} \ellipticK{k}}{S}\;
  \Jacobisnadjustsquareb{\frac{2 s \,\ellipticK{k} }{S}}{k}
\; ,
  \label{eq:kappa}
\end{equation}
where $S:=(2\pi + \se)/2n=\st/2n$.
Closure of the surface in Euclidean space determines the parameter $k$ (see the  discussion following Eq.~(\ref{eq:varphi}) in section~\ref{subsec:zeromodes}).
It is directly related to the stress $C_{\|}$ and the constant of integration $J$ via
$C_{\|}=1 - 4 (1+k)\ellipticK{k}^{2}/S^{2}$ and
$J=\sqrt{C_{\|}^{2} -64 \,k\ellipticK{k}^{4}/S^{4}}$.
In Eq.(\ref{eq:kappa}) the function $\Jacobisnadjustroundb{s}{k}$ is the sine of the Jacobi
amplitude $\Jacobiamadjust{s}{k}$ with parameter $k$. The symbol $\ellipticK{k}$
denotes the complete elliptic integral of the first kind \cite{Abramowitz}.


\subsection{\sf Deformed cones}

To assess the stability of the cones with surplus angle, we need to examine the second variation of the energy.  Unlike the unique ice-cream cone, the question of stability is non-trivial. For technical simplicity we will limit the discussion to cones which do not exhibit self-contact, $i.e.$ $\se<7.08$ for $n=2$, for example \cite{econe}.

\vskip1pc\noindent  The local deformations of the cone
will not only be constrained by isometry they must also preserve
its apex. The deformed cone is thus another cone.
This can be described in terms of the
deformation of the trajectory on the unit sphere, $\VECu(s)\to
\VECu(s)+ \delta \VECu(s)$. The only non-vanishing component of the deformed
extrinsic curvature tensor is then given by $\delta_\VECX K_{ss} =  r \delta_\VECu \kappa$.
One thus only needs to evaluate $\delta_\VECu \kappa$.

\vskip1pc\noindent The unstretchability of the cone
translates into the statement of conservation of arc-length on
this sphere. In particular, the angle surplus is preserved. The
change in arc-length induced by $\delta \VECu$ is given by
$\delta ds = \VECt\cdot (\delta \VECu)'\, ds$. Thus fixed arc-length is captured by the identity
\begin{equation}
\VECt\cdot (\delta \VECu)'=0\,. \label{eq:arclfix}
\end{equation}
It is useful to recast Eq.(\ref{eq:arclfix}) as a relationship
connecting the two non-vanishing projections of $\delta \VECu$,
$\Phi= \VECn\cdot\delta\VECu$ and $\Psi=\VECt\cdot \delta
\VECu$:
\begin{equation}
\Psi ' = - \kappa\, \Phi \,. \label{eq:arccon}
\end{equation}
This is a special case of Eq.(\ref{isomgen}) with the identification
$\phi=r\Phi$, $\psi_a = r  \Psi \VECt\cdot \VECe_a$.
In particular, note that unless $\kappa$ vanishes, every
non-vanishing normal deformation will be accompanied by a
compensating tangential deformation. In general,
a global constraint on $\Phi$ is implied by Eq.(\ref{eq:arccon}), namely
\begin{equation}
\label{eq:arccon1}
\oint ds\, \kappa \Phi \equiv \,\langle \kappa \,|\, \Phi \rangle = 0
\,.
\end{equation}

\vskip1pc
\noindent Because arc-length is conserved, derivatives with
respect to $s$ commute with the variation. Thus, in particular,
\begin{equation}
\delta \VECt= (\delta \VECu)'\,.
\end{equation}
$\delta \VECn$
 is thus expressed in terms of $\delta \VECu$
as follows
\begin{eqnarray}
\delta \VECn &=& - (\VECn\cdot \delta \VECu)\, \VECu -
(\VECn\cdot \delta \VECt)\, \VECt\nonumber\\
&=& - (\VECn\cdot \delta \VECu)\, \VECu - (\VECn\cdot
\delta \VECu')\, \VECt\,.
\end{eqnarray}
We are also now in a position to
write down how the curvature $\kappa= - \VECn\cdot
\VECt'$ changes. We have, for a length preserving deformation,
\begin{eqnarray}
\delta_\VECu \kappa &=& - \VECn\cdot (\delta \VECt)' - \delta {\VECn}
\cdot \VECt'\nonumber\\
&=& -\VECn\cdot [\delta \VECu + \delta \VECu'']\,.
\end{eqnarray}
The second structure equation appearing in (\ref{frameprime}) was
used on the second line.  Casting $\delta \kappa$
in terms of the projections, $\Phi$ and $\Psi$, we obtain
\begin{equation}
\delta \kappa = - \Phi'' - \Phi + [\kappa \Psi]'\,.
\label{eq:delkcomp}
\end{equation}
This equation is a special case of Eq.(\ref{delK2}).


\subsection{\sf Second Variation}

The second variation (\ref{eq:del2HX}) of the energy now reads, modulo a constant multiplicative factor originating in the integration with respect to r,
\begin{equation}
\delta^2 H= - \oint ds \, \Phi\, [\delta \kappa'' +
\left(\frac{3}{2}\kappa^2 + 1-C_\|\right) \delta \kappa] \,. \label{eq:del2H}
\end{equation}
This is also what one would have guessed without going through a detailed
 analysis.  The naive term linear in the variation of the multiplier
$\delta C_\|$ (assumed to be constant) vanishes on account of Eq.(\ref{eq:arccon1}).  The
justification for ignoring $\delta C_\|$ is found in section~\ref{2ndvar}.
Modulo this caveat, the calculation adapted directly to the conical geometry
does get the answer right.

\vskip1pc\noindent
Let us now substitute the expression for $\delta \kappa$
given by Eq.(\ref{eq:delkcomp}) into Eq.(\ref{eq:del2H}).
One finds that the dependence on the tangential deformation $\Psi$ can be collected in a total derivative so that
the second variation can be expressed completely in terms of $\Phi$:
\begin{equation} \label{eq:del2HVs}
\delta^2 H = \oint ds \, \Phi \Big\{\Phi'''' + V_1(s)\Phi '' + V_2(s) \Phi \Big\}\,,
\end{equation}
where
\begin{eqnarray}
V_1(s) &=& \frac{5}{2}\kappa^2 + 2- C_\|\nonumber\\
V_2(s) &=& \frac{1}{2} \kappa\kappa'' + \frac{1}{2}(\kappa')^2 + \kappa^4 +\frac{3}{2}\kappa^2 +
1-C_\| \nonumber\\
& = & \frac{5}{8}\kappa^{4} + [\frac{3}{2} - (1-C_{\|})] \kappa^{2}
 + \, \frac{1}{2}(J^{2} - C_{\|}^{2}) + 1-C_{\|}   \,.
\end{eqnarray}
The second presentation of $V_2$ uses the Euler-Lagrange equation as well as the quadrature~(\ref{eq:firstintegral}).
The details of this derivation are collected in appendix \ref{app:2ndvarcone}.

\vskip1pc\noindent
The obtained expression is still not in a manifestly self-adjoint form. Rearranging
derivatives
\begin{equation}
\Phi (V_1 \Phi')' = V_1 \Phi\Phi '' + \frac{1}{2} \, V_1' (\Phi^2)'\,,
\end{equation}
permits Eq.(\ref{eq:del2HVs}) to be cast in the form
\begin{equation}
\delta^2 H = \oint ds \, \Phi \Big\{\Phi'''' + [V_1(s) \Phi']' +
\left(V_2(s)+ \frac{1}{ 2} V_1''(s)\right) \Phi \Big\}\,.
\label{eq:del2Hsa}
\end{equation}
We write
\begin{equation}
\delta^2 H = \oint ds \, \Phi {\cal L} \Phi \equiv \,\langle \Phi\,|\,{\cal L} \,|\, \Phi \rangle
\,,
\end{equation}
where the fourth order linear differential
operator ${\cal L}$,  given by
\begin{equation} {\cal L}=
{\partial^4\over \partial s^4} + {\partial\over \partial s}\,
V_1(s) {\partial\over \partial s}  + V_2(s)+ \frac{1}{2} V_1''(s) \,,
\label{eq:operatorL}
\end{equation}
is self-adjoint.  As such,  ${\cal L}$  is guaranteed to possess real eigenvalues.
If all these eigenvalues are
positive (for a given $n$ and $\se$), the corresponding equilibrium shape will be
stable.


\section{\sf Stability}

\subsection{\sf Small Surpluses\label{subsec:smallsurpluses}}

\begin{figure}
\begin{center}
 \includegraphics[width=0.6\textwidth]{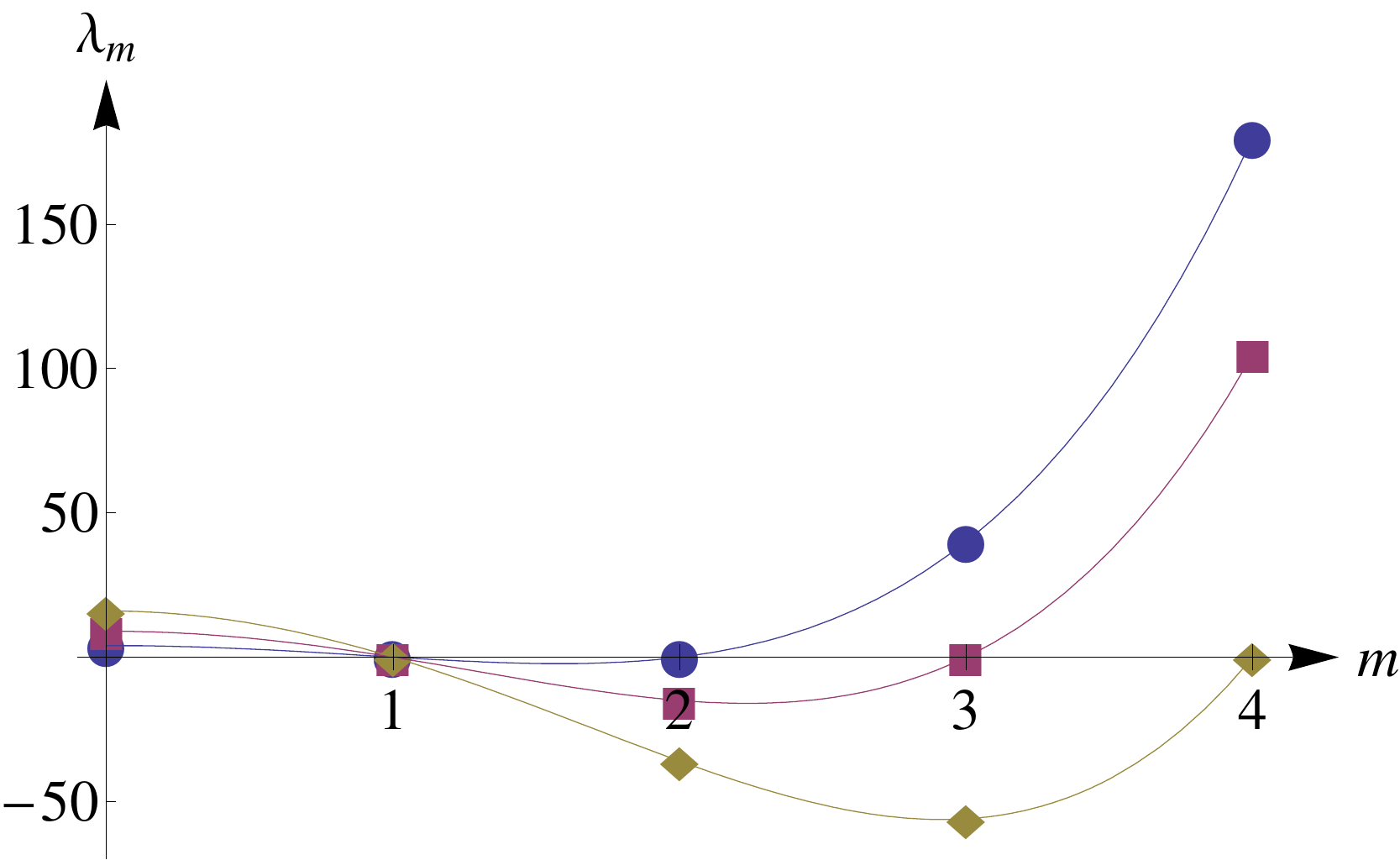}
\end{center}
\caption{Eigenvalues $\lambda_m$ in the small surplus limit as a function of mode number $m$ for
$n=2$ (circles), $n=3$ (squares), and $n=4$ (diamonds). The joining lines correspond to the
eigenvalue Eq.(\ref{eq:eigenvaluessmallsurplus}) but have no physical significance for non-integer
$m$; they have been added as a guide to the eye. Note that for $m\ge 1$ each point corresponds to
two modes due to the two-fold degeneracy of the small
surplus limit.}\label{fig:eigenvaluessmallsurplus}
\end{figure}

Before analyzing the spectrum of ${\cal L}$ in the full non-linear theory, we provide a complete
stability analysis of cones with a small surplus angle $\se$. In this case $\kappa$ is also small,
so that the quadrature~(\ref{eq:firstintegral}) implies that the equilibrium curvature is given by
$\kappa= \kappa_0 \sin (n s)$  where $n\ge 2 \in \mathbb{N}$ (we fix the phase without loss of generality) which
identifies $C_\| = 1 - n^2$ and $J^2 = C_\|^2$ to lowest order. The ground state corresponds to $n=2$.
In this approximation, both $V_1$ and $V_2$ are constants,  $V_1= n^2 + 1 $ and $V_2 = n^2$ so that
\begin{equation} {\cal L}=
{\partial^4\over \partial s^4} + (1 + n^2) \,
 {\partial^2\over \partial s^2}  + n^2\,.
\end{equation}
Periodicity implies that the modes of deformation are represented by a constant ($m=0$),  $\sin (m s)$ and $\cos (m s)$,
$m\ge 1\in \mathbb{N}$.
For fixed $n$, the eigenvalues of ${\cal L}$ are then labeled by the integer $m=0,1,2,\dots$, given
by
\begin{equation}
\lambda_m = (m^2- n^2)( m^2-1) \,.
\label{eq:eigenvaluessmallsurplus}
\end{equation}
In Fig.~\ref{fig:eigenvaluessmallsurplus} $\lambda_m$ is plotted as a function of $m$ for the lowest values of $n$.
The constant mode with $m=0$ has positive $\lambda_0= n^2$.
All eigenvalues with $m\ge 1$ possess a two-fold degeneracy.
As we will see this degeneracy gets lifted for some modes when
the angle surplus is increased.

\vskip1pc\noindent
Four zero modes with $\lambda_m =0$
occur at $m=1$ and $m= n$.
The mode $\cos (n s) \propto \kappa'$ describes a small rotation of the cone about the axis of
$n$-fold symmetry $\VECk$.  The two modes with $m= 1$ describe rotations about an orthogonal  axis.
Thus these three modes are the zero modes predicted by the rotational invariance of the conical bending energy. The fourth zero mode $\sin (n s)$ is inconsistent with the isometry constraint (\ref{eq:arccon1}). It is also the \emph{only} conical mode of deformation inconsistent with isometry in the small surplus limit.
If isometry is enforced, the mode does not persist as an eigenmode. As we will see in
section~\ref{subsec:Fourierdecomp}, if the isometric constraint is relaxed it becomes unstable for
higher surplus angles.

\vskip1pc\noindent
The 2-fold ground state with $n=2$ is stable. All excited states of the cone ($n\ge 3$) are unstable
for small surpluses. This is contrary to what was claimed in Ref.~\cite{econe}. In total $2(n-2)$ unstable
modes of deformation exist corresponding to $m=2, \dots, n -1$. They lie between the zero modes at
$m=1$ and $m=n$ (see Fig.~\ref{fig:eigenvaluessmallsurplus}). All modes of deformations with $m > n$ make a positive contribution to the
energy.

\vskip1pc\noindent
The dominant (fastest) modes of instability will be the two modes corresponding to the lowest
negative eigenvalue. The corresponding value of $m$ is the integer closest  to
$m_0=\sqrt{(1+ n^2 )/2}$.
For $n=3$, the only possibility is $m=2$;
for $n=4$, the $m= 3$ mode has the lower eigenvalue.
For higher values of $n$,  one would expect a cascade of instabilities. For example, the instability associated with the state with $n=7$ will proceed by
$7\to 5 \to 4 \to 3\to 2$, at least for small $\se$.
However, predicting the dynamics of this cascade
is beyond the scope of this analysis.
The final part of this work will focus on the
extent to which the results of this section hold outside the small surplus limit.


\subsection{\sf Zero modes in the non-linear regime\label{subsec:zeromodes}}

\noindent  One can use the symmetry of the system to
identify the zero modes of the operator ${\cal L}$
satisfying the linearized Euler-Lagrange equation
\begin{equation}
\label{eq:phi0}
{\cal L} \, \Phi =0\,.
\end{equation}
As mentioned above, three zero modes are anticipated by the invariance of the conical bending
energy with respect to rotations about its apex.  Consider the infinitesimal rotation $\delta
\VECu= \VECu\times \VECb$.
Its normal projection, given by  $\Phi_{\VECb}= {\VECb}\cdot \VECt$,
then satisfies Eq.(\ref{eq:phi0}) for each choice of ${\VECb}$.\footnote{Note that $\Phi_{\VECb}$ is not a solution of
the non-self-adjoint equation
\begin{equation}
\Phi'''' + V_1(s)\Phi '' + V_2(s) \Phi =0 \,.
\end{equation}
While the total energy does vanish, the integrand
appearing in Eq.(\ref{eq:del2HVs}) does not vanish point wise.
}
While the quadratic constraint $\VECt^2=1$
connects the three components, this is not relevant
at the linear level.

\vskip1pc\noindent It is simple to check that these three modes are, in fact, consistent with global isometry:
\begin{equation}
\langle \kappa | \Phi_{\VECb} \rangle = \VECb \cdot \oint ds\, \kappa \VECt
 \stackrel{(\ref{frameprime})}{=} \VECb \cdot \oint ds \,\VECn'=0\,.
\end{equation}
In our analysis of small $\se$ a fourth zero mode proportional to $\kappa$ was identified.
However, this mode of deformation was seen to be inconsistent with the global
isometry constraint and can thus be discarded here.

\begin{figure}
\begin{center}
  \includegraphics[width=0.47\textwidth]{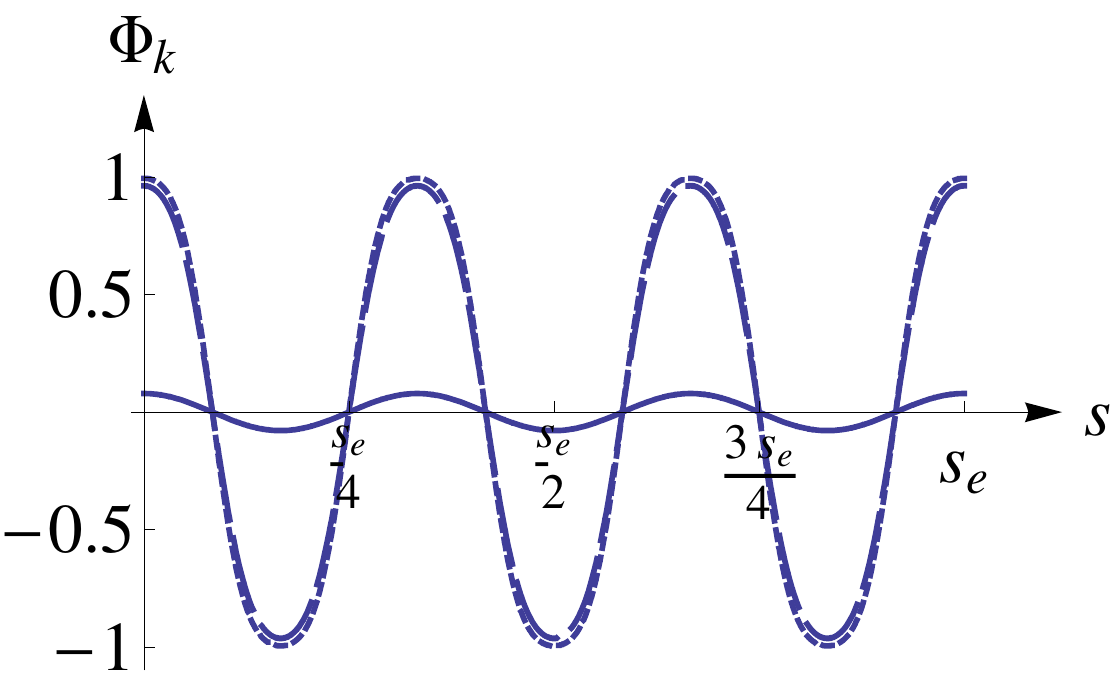}
\end{center}
\caption{The zero mode $\Phi_{\VECk}$ for $n=3$ and $\se=\frac{\pi}{360}$ (solid line),  $\pi$
(dashed line), and $2\pi$ (dotted line).}\label{fig:1stzeromode}
\end{figure}
\vskip1pc\noindent
The three zero modes can be expressed as projections of $\VECt$ onto the Euclidean
coordinate system $(\VECi,\VECj,\VECk)$. Using the fact that $\VECk$ is aligned with $\VECJ$ we
obtain with the help of Eq.(\ref{eq:VECJ}):
\begin{equation}
  \Phi_{\VECk} = \VECk \cdot \VECt = \kappa'/J \; .
\end{equation}
which identifies the first zero mode with $\kappa'$ up to a normalization (see
Fig.~\ref{fig:1stzeromode} for a numerical example with $n=3$).
However, we also know that $\kappa'$ satisfies the linearized Euler-Lagrange equation
(\ref{eq:k3p}),
\begin{equation}
\delta \kappa''  +\left( \frac{3}{2}\kappa^2 + 1-C_\|\right)\delta \kappa =0\,.
\label{eq:linel}
\end{equation}
This is a direct consequence of the reparametrization invariance of the
Euler-Lagrange equation and is due to the fact that a rotation about
the axis of symmetry can be achieved by a reparametrization.

\begin{figure}
\begin{center}
  \subfigure[]{\includegraphics[width=0.47\textwidth]{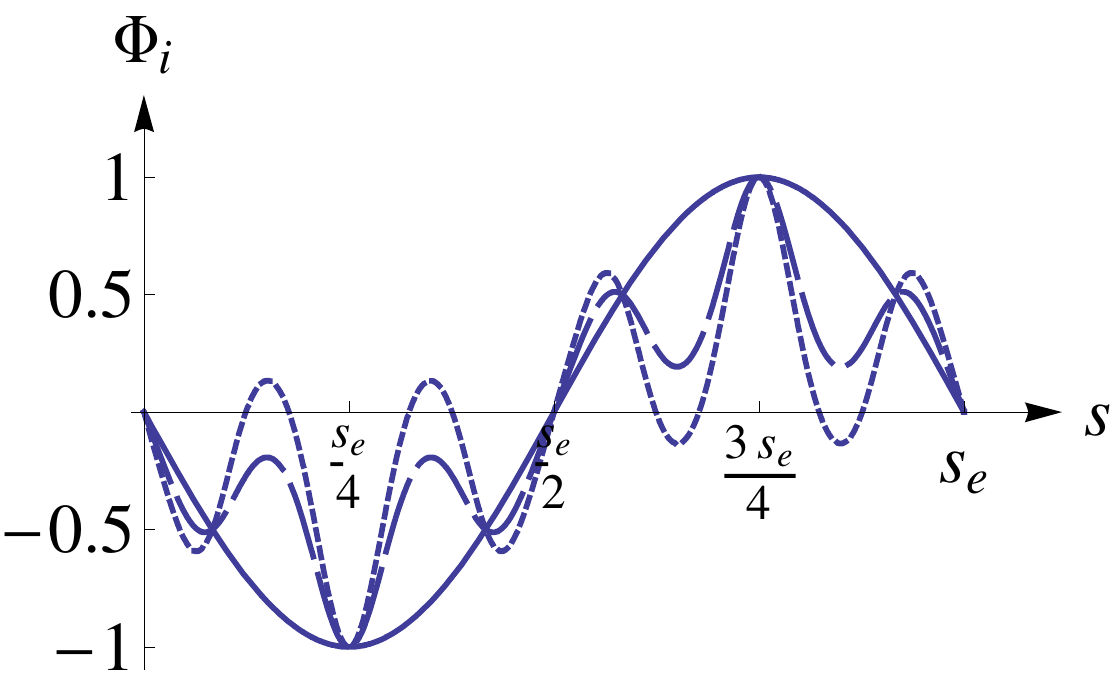}}
  \hfill
  \subfigure[]{\includegraphics[width=0.47\textwidth]{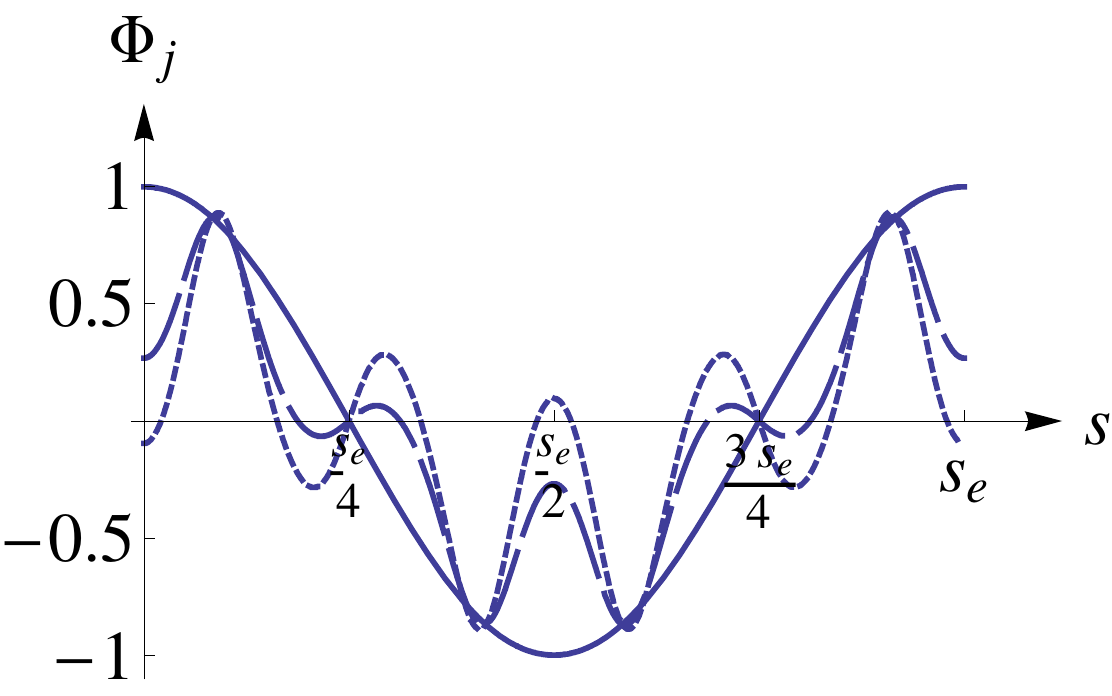}}
\end{center}
\caption{The zero modes $\Phi_{\VECi}$ and $\Phi_{\VECj}$ for $n=3$ and $\se=\frac{\pi}{360}$
(solid line),  $\pi$ (dashed line), and $2\pi$ (dotted line).}\label{fig:2nd3rdzeromodes}
\end{figure}
\vskip1pc\noindent
The second and third zero modes are more complicated and their construction requires a little more work.
We start by expressing the vector $\VECu=\cos\varphi\sin\vartheta\,\VECi
+\sin\varphi\sin\vartheta \,\VECj +\cos\vartheta\,\VECk$ in terms of the polar and azimuthal angles
on the sphere, $\varphi$ and $\vartheta$ respectively. Both angles are functions of arc-length $s$.
With the help of the first structure equation appearing in (\ref{frameprime}) it is possible to write
\begin{equation}
  \Phi_{\VECi}  = \VECi \cdot \VECu' = - z_1 \sin{\varphi} + z_2 \cos{\varphi} \quad \text{and} \quad
  \Phi_{\VECj}  = \VECj \cdot \VECu' = z_1 \cos{\varphi} + z_2 \sin{\varphi}
  \label{eq:2nd3rdzeromode}
\end{equation}
with $z_1=\varphi' \sin{\vartheta}$ and $z_2=\vartheta'\cos\vartheta$. From the projection onto $\VECk$, $\Phi_{\VECk}$, one identifies
$\vartheta' = -\kappa'/(J \sin\vartheta)$.
Using the identity $\cos{\vartheta} = \VECu\cdot\VECk \stackrel{(\ref{eq:VECJ})}{=} \kappa/J$,
the $\vartheta$ dependence in the zero modes can be eliminated to obtain
\begin{equation}
  z_1 = \varphi' \, \sqrt{1-\frac{\kappa^2}{J^2}} \quad \text{and} \quad z_2 = - \frac{\kappa \kappa'}{J \sqrt{J^2-\kappa^2} }
  \; .
  \label{eq:z1z2}
\end{equation}
To replace the $\varphi$ dependence, consider the projection $\VECn\cdot\VECk =
(\VECu\times\VECu')\cdot\VECk = z_1 \sin{\vartheta} \stackrel{(\ref{eq:VECJ})}{=}(\kappa^2/2 -
C_\|)/J$.
From this it follows that $\varphi'=(\kappa^2/2 - C_\|)/[J(1-\kappa^2/J^2)]=(J/2)\{[(J^2-2C_\|)/(J^2-\kappa^2)]-1\}$. An integration with respect to the arc-length completes the elimination of
the angular dependences in Eq.(\ref{eq:2nd3rdzeromode}). One obtains
\begin{equation}
\varphi (s) = \frac{J^{2} - 2C_{\|}}{2 J}
    \frac{S}{2 \ellipticK{k}}
    \operatorname{\Pi}\left[\frac{16(-k)\ellipticK{k}^{2}}{J^{2} S^{2}},
    \Jacobiamadjust{\frac{2 \ellipticK{k} \, s}{S}}{k},k\right]
    - \frac{J}{2} s
  \; ,
\label{eq:varphi}
\end{equation}
where $\Pi$ is the elliptic integral of the third kind and $\Jacobiamadjust{s}{k}$is the Jacobi amplitude with parameter $k$ which appears in the expression~(\ref{eq:kappa}) for the curvature $\kappa (s)$.
The closure of the surface in $\mathbb{R}^3$ implies that $\varphi (\st)=2\pi$. From this condition the parameter $k$  can be determined.

\vskip1pc\noindent
In Fig.~\ref{fig:2nd3rdzeromodes} numerical examples of $\Phi_{\VECi}$ and $\Phi_{\VECj}$ are shown
for $n=3$. For small $\se$ the modes are proportional to $\sin{s}$ and $\cos{s}$ as anticipated in
the small surplus analysis of the previous section.  This behavior changes significantly as soon as
the non-linearity kicks in: higher frequencies blend in and further nodes appear which complicates
the identification of zero modes from their functional dependence. Note that this problem does not
arise with $\Phi_{\VECk}$. Being proportional to $\kappa'$, it always possesses $2n$ nodes no matter
what value of $\se$ is given (see again Fig.~\ref{fig:1stzeromode}).

\vskip1pc\noindent
The nodal structure of the zero modes suggests one of the difficulties that
will be encountered in the full perturbation analysis: a simple identification of the number of
nodes with an integer $m$ in the manner of the small surplus limit will become unfeasible. How one
sidesteps this obstacle
will be discussed in the Fourier analysis in section \ref{subsec:Fourierdecomp}.
Before turning to the full problem let us have another look at the constant mode.


\subsection{\sf Constant normal deformations}

For small $\se$, we saw that the constant normal deformation represents a stable eigenmode of
deformation with $m=0$ and $\lambda_0= n^2$. The constant mode will not generally be an eigenmode.
However, it is possible to show that $\delta^2 H$ is positive for such a deformation.

\vskip1pc \noindent
Rewrite Eq.(\ref{eq:del2Hsa}) in the form
\begin{equation}
\delta^2 H = \oint ds \, \Big\{(\Phi'')^2 - V_1(s) (\Phi')^2 + \left(V_2(s)+ \frac{1}{2}
V_1'' (s) \right) \Phi^2  \Big\}\,.
\label{eq:del2Hsq}
\end{equation}
The second term appearing in Eq.(\ref{eq:del2Hsq}) is manifestly negative.
If $\Phi$ is constant, then \footnote {\sf Such a deformation is consistent with
Eq.(\ref{eq:arccon1}). The identity $\oint ds \kappa =0$
follows from the symmetry of the quartic potential.}
\begin{equation}
\delta^2 H= \Phi^2 \,\oint ds \, V_2 (s)\,.
\end{equation}
Modulo the Euler-Lagrange equation,
\begin{equation}
V_2(s) = \frac{1}{2}(\kappa')^2 + \frac{3}{4}\kappa^4 + \left(1 + \frac{1}{2}C_\|\right) \kappa^2 +
(1-C_\|) \,,\end{equation}
which is manifestly positive for each $n$ for values of $C_\|$ in the range $1-n^2 \le C_\| \le 1$. Outside this range $V_2$ can assume negative values.
However, we have confirmed numerically for the lowest $n$-folds that the integral itself is indeed positive for all $\se$ lying outside the regime where self-contacts occur; a transition from one $n$-fold to another cannot be induced by a constant normal deformation.

\vskip1pc\noindent
Now let us look more generally at the spectrum of ${\cal L}$ in the full non-linear theory.
Let ${\cal L} \Phi_i = \lambda_i\Phi_i$, $i=1,2,3,\dots$ ordered
by the magnitude of $\lambda_i$ with periodic boundary conditions on $\Phi_i$ and ${\cal L}$ given by Eq.~(\ref{eq:operatorL}).
In physics one is perhaps more familiar with second order differential operators of the form
$-\partial_s^2 + V(s)$.  In this case the number of nodes in the eigenmodes increases with energy.
We have seen that the behavior of the fourth-order operator ${\cal L}$ is very different. While the fourth derivative term
will always dominate if the mode oscillates sufficiently, as we have seen, for modes commensurate
with the $n$-fold,  the negative second derivative term can lower the mode energy.   It is this term
which conspires to produce zero modes exhibiting $2$ and $2n$ nodes if the surplus angle is small.
For all $n>2$, it is not unreasonable to expect that  unstable modes exist in the nonlinear theory
as well corresponding to those with $4,6,\dots, 2(n-1)$ nodes in the small surplus limit (see
section~\ref{subsec:smallsurpluses}).


\subsection{\sf Fourier decomposition of modes\label{subsec:Fourierdecomp}}

The principal technical obstacle to analyzing the
stability of a cone is
the non-local nature of the isometry constraint: the eigenmodes are orthogonal
to $\kappa$ as vectors in a Hilbert space. The constraint makes it unlikely that one is going to
make further progress without recourse to some form of approximation. Our approach to the problem
will make use of techniques of perturbation theory in quantum mechanics. The language and notation
we adopt reflects this point of view.

\vskip1pc\noindent
We first define an orthonormal  basis of periodic functions $\{\tilde{\phi}_i\}$, $i=1,\ldots,i_{\text{max}}$,
that is orthogonal to $\kappa$. These will be constructed out of
Fourier modes using the Gram-Schmidt process.
We begin with the Fourier modes $\phi_1 = 1$, $\phi_{2m} = \sin{\left(m \frac{2\pi s}{\st} \right)}$, and
$\phi_{2m+1} =\cos{\left(m \frac{2\pi s}{\st} \right)}$ for $m=1,\ldots,m_{\text{max}}$, where $\st=2\pi + \se$ as before.
We normalize them with respect to the scalar product defined on the Hilbert space,
$\hat{\phi}_{i} := \sqrt{\frac{2-\delta_{i1}}{\st}} \, \phi_{i}$.
Certain Fourier modes will not, however, be
consistent with the constraint
$\langle \kappa|\hat{\phi}_i \rangle =0$.
To see this, let us expand expression~(\ref{eq:kappa}) for $\kappa (s)$ into a Fourier series:
\begin{equation}
  \kappa (s) = \frac{16\pi n}{\st} \sum_{l=0}^{\infty} (-1)^{l} \frac{q^{l+\frac{1}{2}}}{1+ q^{2l+1}}
  \sin{\left[(2l+1) n \, \frac{2\pi s}{\st} \right]}
  \; ,
  \label{eq:kappaFourierseries}
\end{equation}
where
\begin{equation}
  q := \text{exp}{\left(- \frac{\pi \,\ellipticK{1/k}}{\sqrt{-k}\,\ellipticK{k}} \right)}
  \; .
\end{equation}
Thus, the sine functions with $m= (2l +1)n$ and $l\in\{0,1,\ldots\}$ are inconsistent with the isometry constraint:
\begin{eqnarray}
  \langle \kappa\,|\, \phi_{2(2l+1)n} \rangle
  & = & \int_{0}^{\st } \! ds \; \kappa \, \sin{\left[ (2l+1) n\, \frac{2\pi s}{\st} \right]}
  = 8 \pi n (-1)^{l} \frac{q^{l+\frac{1}{2}}}{1+ q^{2l+1}}
  \nonumber \\
  & = &
  \frac{4 \pi n (-1)^{l}}{\cosh{\left[ \frac{\pi \,\ellipticK{1/k} (l+\frac{1}{2}) }{\sqrt{-k}\,\ellipticK{k}} \right]}}
  \ne 0
  \; .
\end{eqnarray}
We immediately note a distinction between even and odd modes of deformation of the cone.
To construct an orthonormal basis consistent with the
constraint we use the Gram-Schmidt process on the offending modes,
defining the new states for $i=2(2l+1)n$:
\begin{equation}
  \bar{\phi}_{i} := \phi_{i} - \frac{\langle\kappa \,|\, \phi_{i}\rangle}{\langle\kappa \,|\, \kappa\rangle} \, \kappa
  - \sum_{i'<i} \frac{\langle \bar{\phi}_{i'} \,|\, \phi_{i}\rangle}{\langle\bar{\phi}_{i'}\,|\,\bar{\phi}_{i'}\rangle}\,
 \bar{\phi}_{i'}
  \;
\end{equation}
and normalizing them with respect to the scalar product on the Hilbert space,
$\tilde{\phi}_{i}:=\bar{\phi}_{i}/\langle\bar{\phi}_{i}\,|\,\bar{\phi}_{i}\rangle$.
For $i\ne 2(2l +1)n$ we can take the normalized Fourier modes $\tilde{\phi}_{i}\equiv\hat{\phi}_{i}$ as defined above.

\vskip1pc\noindent
For the stability analysis we have to determine the eigenvalues and -vectors of the operator ${\cal
L}$ for pairs of ($n,\se$). In the following we consider in particular  $\se=\frac{\pi}{360}$ ,
$\frac{\pi}{2}$,  $\pi$, and $2\pi$ for $n\le4$. An approximation for the corresponding eigenvalues
and -vectors can be found by diagonalizing the symmetric matrix ${\cal L}_{ij}=\langle \tilde
\phi_{i}|{\cal L}|\tilde \phi_{j} \rangle$ numerically for a fixed $m_{\text{max}}$, $i, j \in \{ 1,
\ldots, 2m_{\text{max}}+1 \}$.\footnote{Certain simplifications can be exploited in the numerical
calculation: Note that $\langle \phi_\text{even} |{\cal L} | \phi_\text{odd} \rangle =0$ on account
of the even parity of ${\cal L}$.
To simplify the calculation of the integrals in  ${\cal L}_{ij}$, the  non-derivative term appearing
in (\ref{eq:operatorL}) is implemented as
\begin{equation}
  V_{2} + \frac{1}{2}V_{1}'' = -\frac{5}{4}\kappa^{4} + [\frac{3}{2} - 6 (1-C_{\|})] \kappa^{2}
    + 3(J^{2} - C_{\|}^{2}) + 1-C_{\|}  \; .
\end{equation}
}
At least the dominant modes should be included to obtain quantitative results, even though low
values of $m_{\text{max}}$ should already be sufficient to make  statements about the stability.
This is due to the fact that higher modes will always make a positive contribution to the energy on
account of the leading fourth derivative term in ${\cal L}$.

\begin{table}
\begin{center}
\subtable[$n=2$]{
\begin{tabular}{|c||c|c|c|c|c|}
\hline &&&&& \\[-0.2cm]
m & $\se$ $\to 0$ &  $\frac{\pi}{360}$ & $\frac{\pi}{2}$ & $\pi$ & $2\pi$ \\[0.2cm]
\hline\hline
0  & 4  & 4.0 & 3.16 & 2.43 & 1.41 \\
\hline
1  & 0  &  0 & 0 & 0 & 0 \\
\hline
2  & 0  & 0  & 0 & 0 & 0 \\
\hline
3  & 40  & 39.64 & 9.75 & 3.60 & 1.16 \\
\hline
4  & 180  & 178.75 & 58.85 $|$ 59.46 & 23.97 $|$ 24.68  &  6.08 $|$  6.38 \\
\hline
5  & 504  & 500.80 & 182.27 & 80.45 & 22.74 \\
\hline
6  & 1120  & 1113.2 & 422.8 $|$ 422.95  & 192.46 $|$ 192.63  & 56.67 $|$ 56.73 \\
\hline
\end{tabular}
}

\vspace*{0.5cm}
\subtable[$n=3$]{
\begin{tabular}{|c||c|c|c|c|c|}
\hline &&&&& \\[-0.2cm]
m  & $\se$ $\to 0$ &  $\frac{\pi}{360}$ & $\frac{\pi}{2}$ & $\pi$ & $2\pi$ \\[0.2cm]
\hline\hline
0  & 9  & 8.99 & 7.47 & 6.04 & 3.81 \\
\hline
1  & 0  &  0 & 0 & 0 & 0 \\
\hline
2  & -15  & -15.04 & -19.10 & -13.98 & -4.90\\
\hline
3  & 0  & 0  & 0 & 0 & 0 \\
\hline
4  & 105  & 103.80 & 19.82 & 8.16 & 3.28\\
\hline
5  & 384  & 380.88 & 105.52 & 39.18 & 10.25\\
\hline
6  & 945  & 938.32 & 302.74 $|$ 307.78 & 120.15 $|$ 127.15 & 28.56 $|$ 33.44 \\
\hline
\end{tabular}
}

\vspace*{0.5cm}
\subtable[$n=4$]{
\begin{tabular}{|c||c|c|c|c|c|}
\hline &&&&& \\[-0.2cm]
m & $\se$ $\to 0$ &  $\frac{\pi}{360}$ & $\frac{\pi}{2}$ & $\pi$ & $2\pi$ \\[0.2cm]
\hline\hline
0  & 16  & 15.99  & 13.53 & 11.09 & 7.12 \\
\hline
1  & 0  &  0 & 0 & 0 & 0 \\
\hline
2  & -36  &  -36.03 & -37.75 & -30.91 & -13.87 \\
\hline
3  &  -56 & -56.30  & -85.63 & -59.75 & -19.73 \\
\hline
4  & 0  & 0  & 0 & 0 & 0 \\
\hline
5  & 216  & 213.01  & 34.10 & 14.80 &  6.4 \\
\hline
6  & 700  & 693.5  & 164.17 & 61.57 & 19.99 \\
\hline
\end{tabular}
}
\end{center}
\caption{Eigenvalues corresponding to the modes $m$ of the flat limit for increasing values of $\se$ and $n=2$, 3, and 4. Each $m$ corresponds to two eigenmodes except $m=0$ (the constant mode) and $m=n$ (the third zero mode). The two-fold degeneracy persists except in those cases where two values are given.} \label{tab:eigenvalues}
\end{table}

\vskip1pc\noindent
To be on the safe side we have taken into account the first two offending modes for $\se=\frac{\pi}{360}$ (by setting
$m_{\text{max}}=9$ (for $n=2$), 14 ($n=3$), and 19 ($n=4$)),
whereas one additional offending mode was included for all higher values of $\se$ (by setting
$m_{\text{max}}=13$ (for $n=2$), 20 ($n=3$), and 27 ($n=4$)).
\begin{figure}[h]
\begin{center}
  \subfigure[$m=2$, odd mode]{\includegraphics[width=0.47\textwidth]{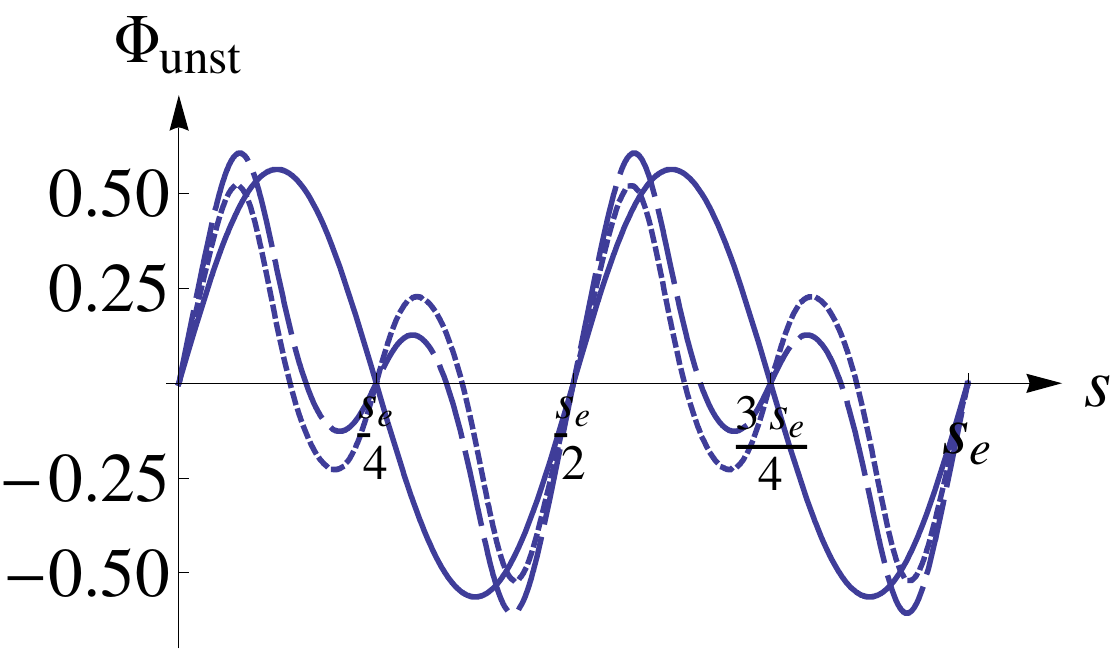}}
  \subfigure[$m=2$, even mode]{\includegraphics[width=0.47\textwidth]{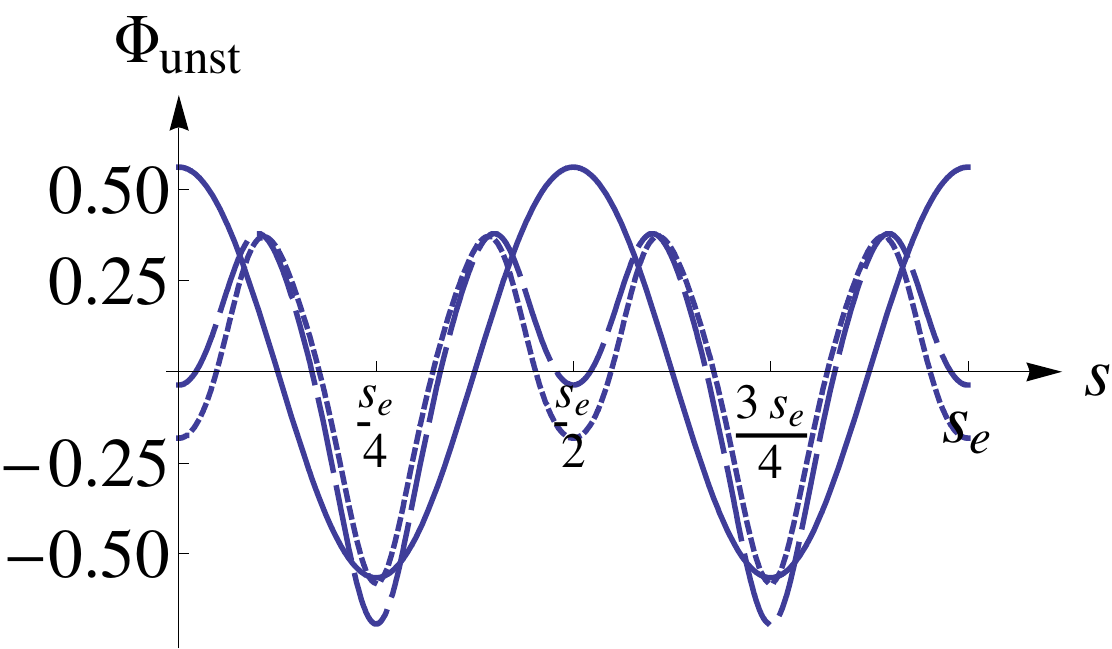}}
\\
\subfigure[Unstable equilibrium state $n=3$]
{\includegraphics[width=0.275\textwidth]{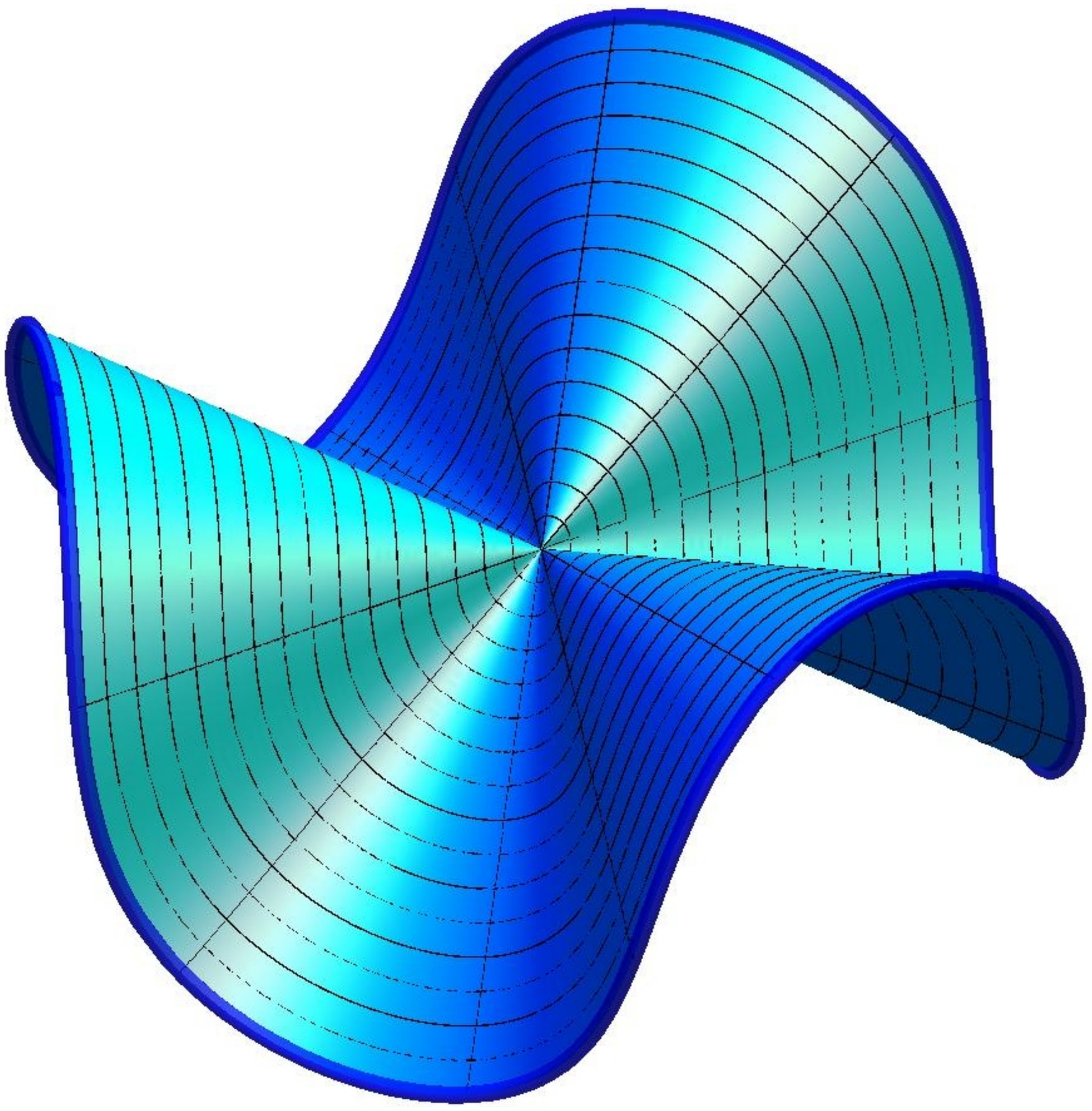}}
\raisebox{3cm}{$\Rightarrow$}
 \subfigure[State $n=3$ deformed with the odd mode $m=2$]
{\includegraphics[width=0.275\textwidth]{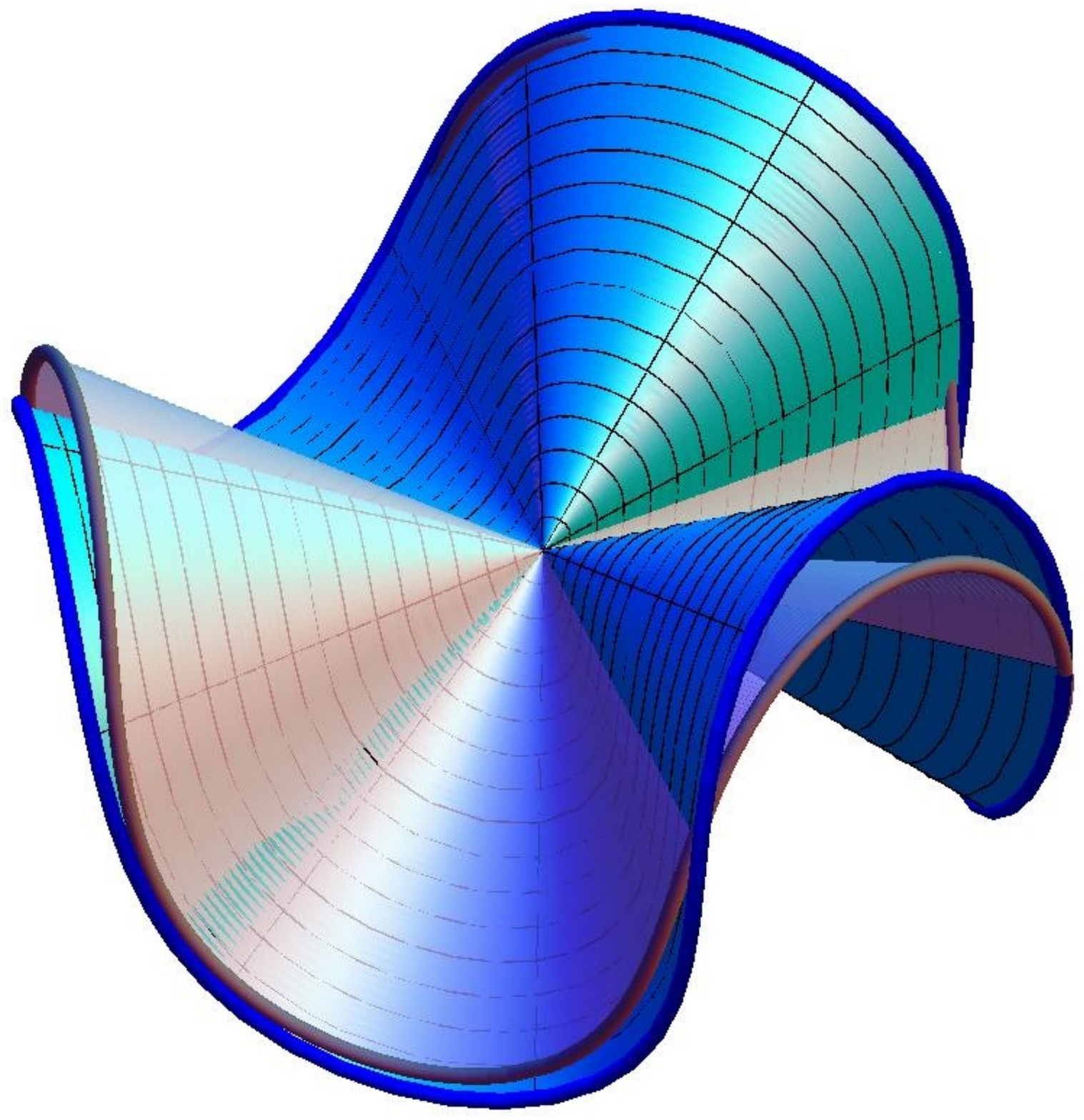}}
\raisebox{3cm}{$\Rightarrow$}
  \subfigure[Stable equilibrium state $n=2$]{\includegraphics[width=0.275
  \textwidth]{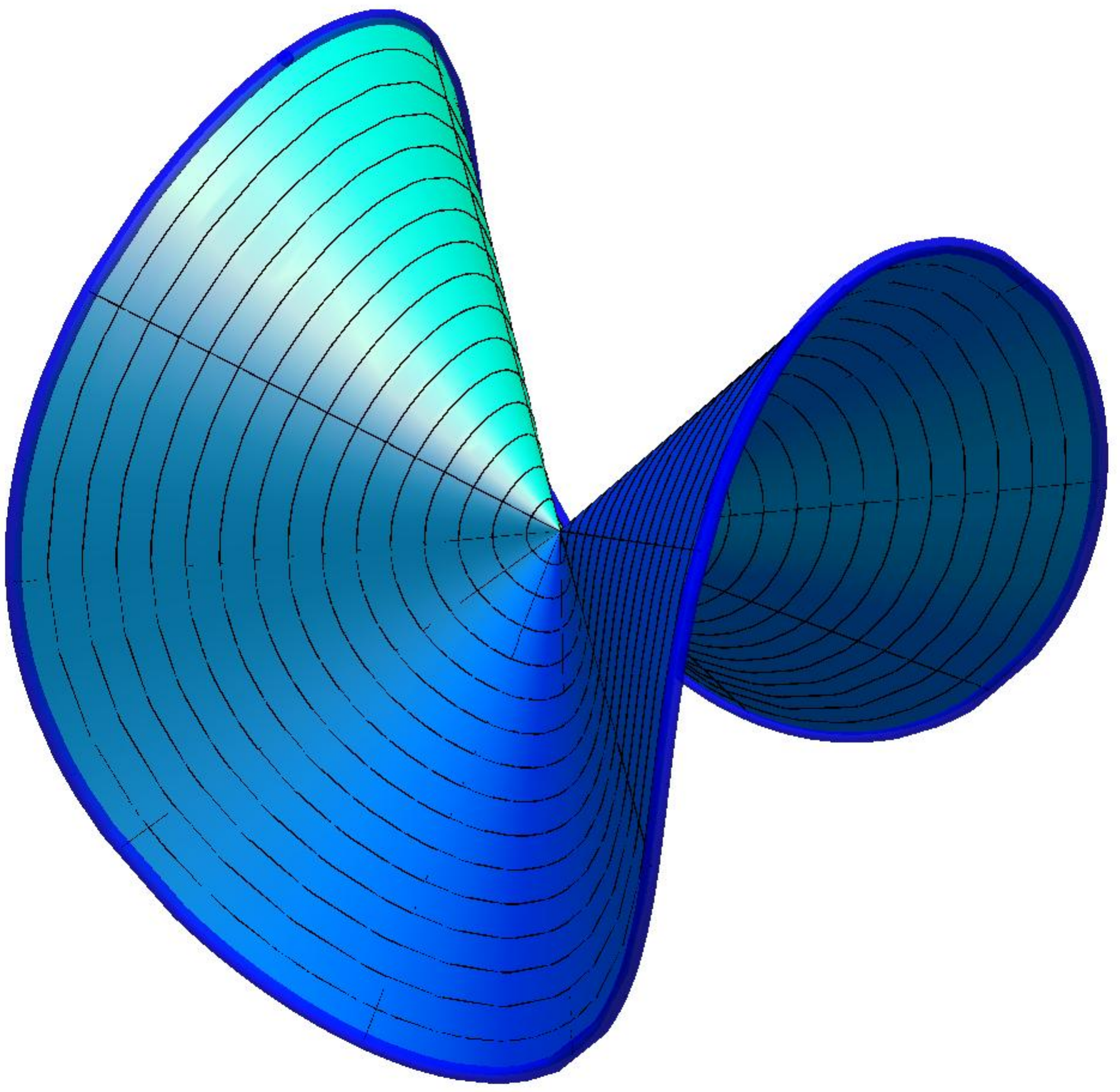}}
\end{center}
\caption{(a) and (b): odd and even unstable modes for $n=3$ corresponding to $m=2$ of the flat
limit for $\se=\frac{\pi}{360}$ (solid line), $\pi$ (dashed line), and $2\pi$ (dotted line).
(c)-(e): transition from the unstable state $n=3$ to the stable state $n=2$ for  $\se=\pi$. In (d) the transparent
cone is the initial state $n=3$ and the cone with the coordinate lines represents its deformation with the odd mode
$m=2$.}  \label{fig:unstablemodesneqthree}
\end{figure}
\vskip1pc\noindent
Tab.~\ref{tab:eigenvalues} presents the lowest eigenvalues for various pairs of $(n,\se)$.
Strikingly, negative eigenvalues are found for $n\ge 3$. A closer inspection
reveals in fact that the stability analysis of the small surplus limit from section~\ref{subsec:smallsurpluses} largely carries over into the non-linear regime: whereas the 2-fold ground state is stable, all excited states of the cone with $n \ge 3$ are unstable against $2(n-2)$ modes of deformation (see Fig.~\ref{fig:unstablemodesneqthree} for the 3-fold). The mode which was dominant in the flat limit continues to be dominant for higher $\se$. This implies that the cascade described at the end of section~\ref{subsec:smallsurpluses} will also occur in the non-linear theory.

\begin{figure}
\begin{center}
  \subfigure[$m=2$, odd mode]{\includegraphics[width=0.47\textwidth]{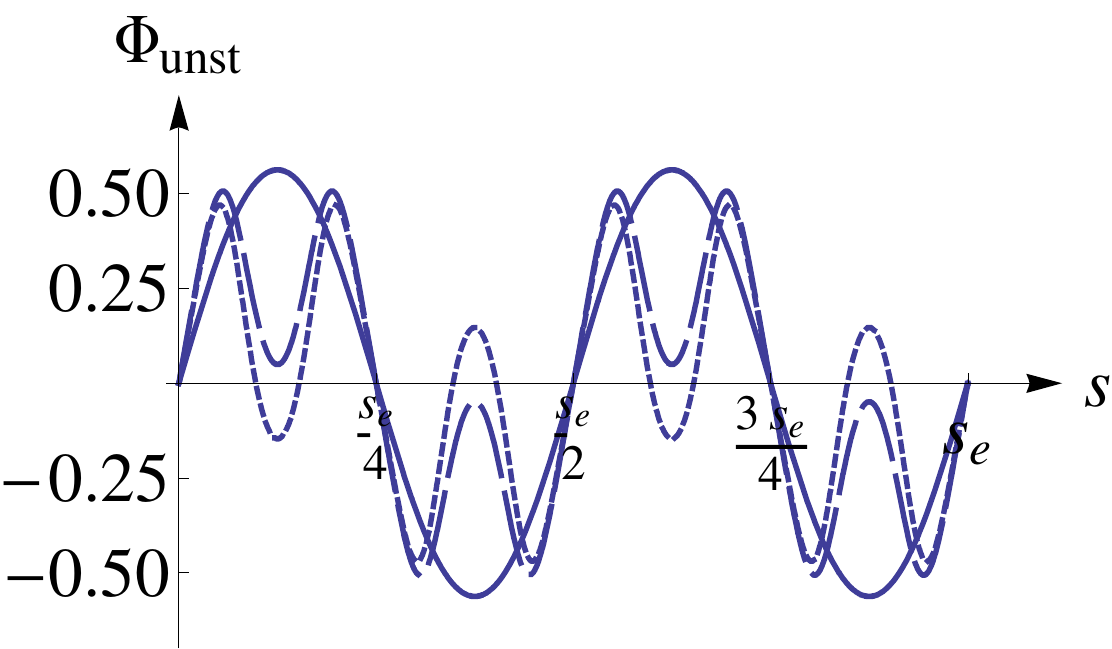}}
  \hfill
  \subfigure[$m=2$, even mode]{\includegraphics[width=0.47\textwidth]{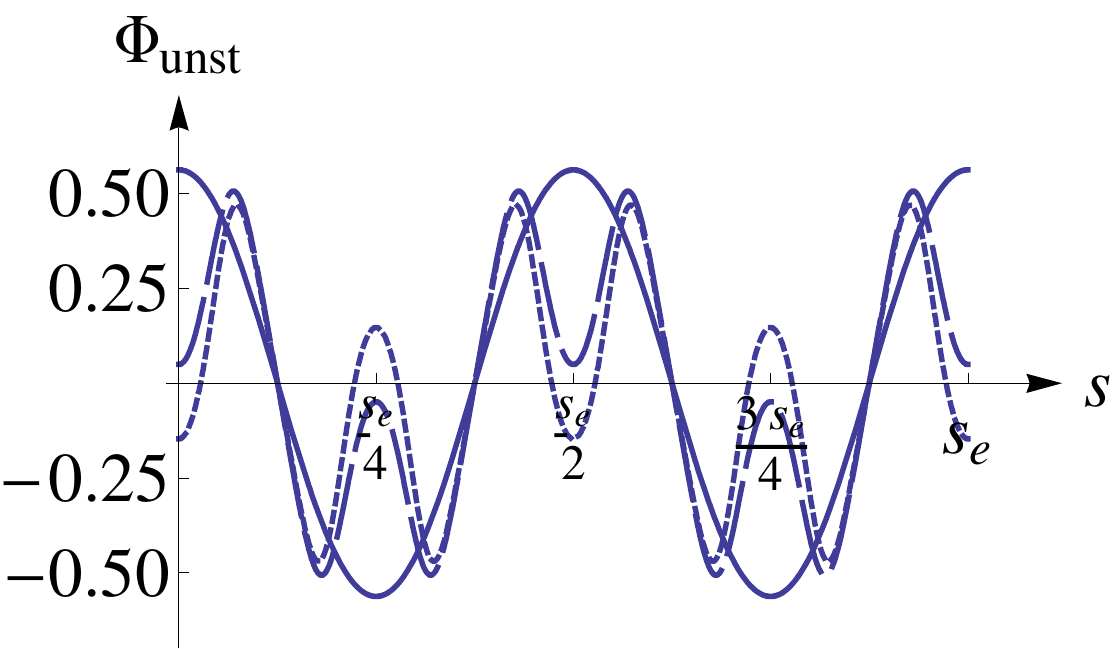}}
\\
  \subfigure[$m=3$, odd mode]{\includegraphics[width=0.47\textwidth]{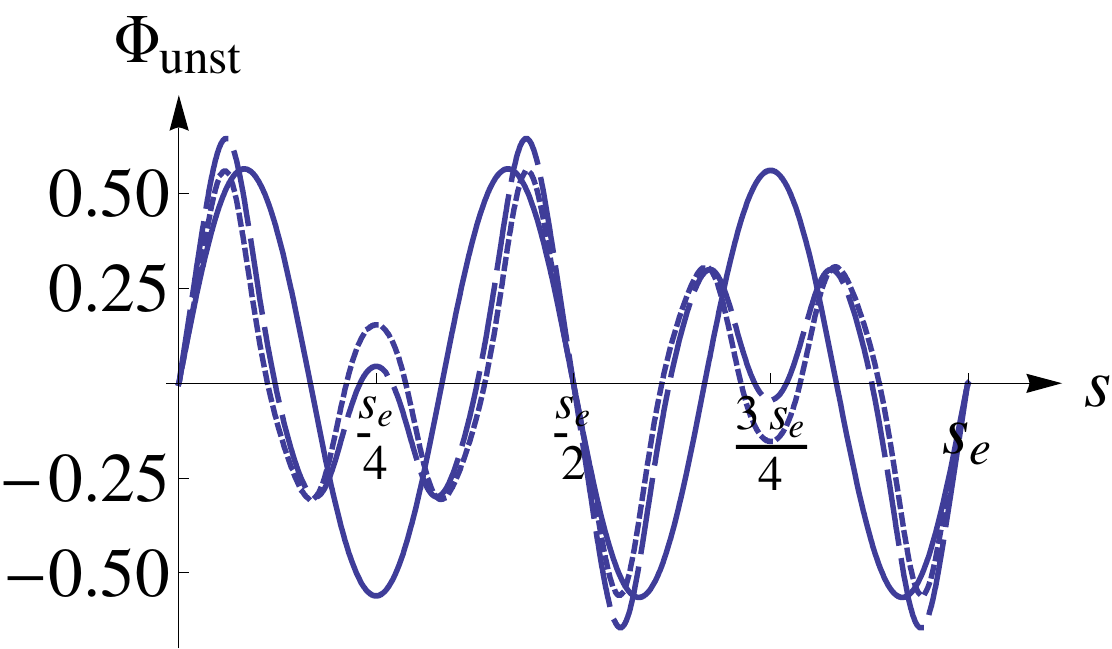}}
  \hfill
  \subfigure[$m=3$, even mode]{\includegraphics[width=0.47\textwidth]{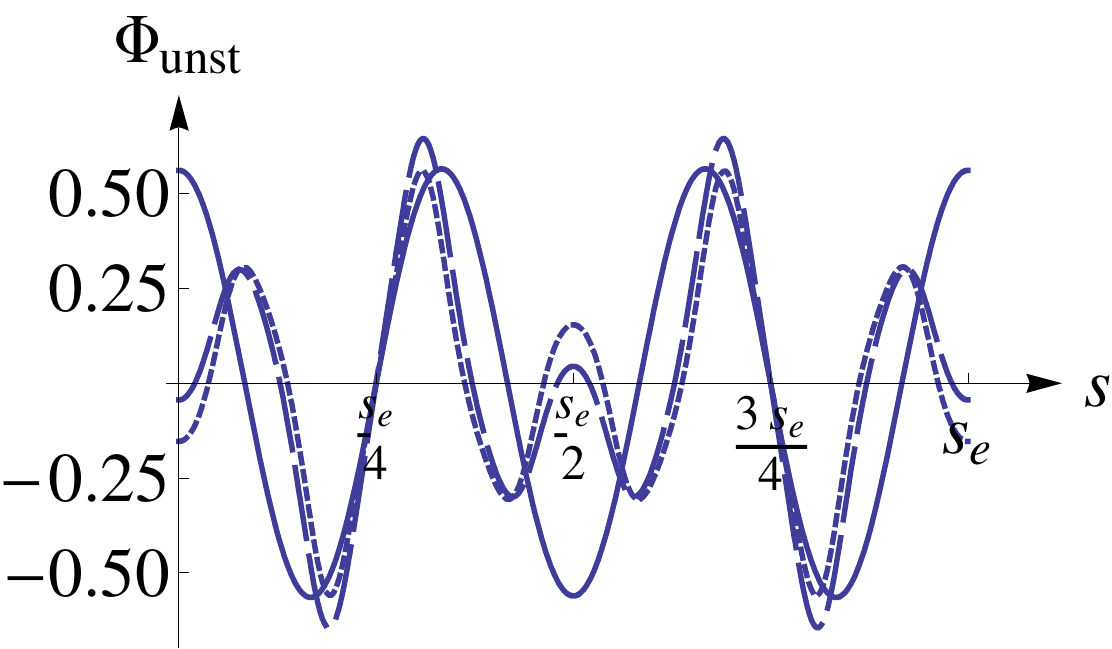}}
\end{center}
\caption{Unstable modes for $n=4$ corresponding to (a), (b) $m=2$ and (c), (d) $m=3$ of the flat
limit for $\se=\frac{\pi}{360}$ (solid line), $\pi$ (dashed line), and $2\pi$ (dotted line).}
\label{fig:unstablemodes}
\end{figure}

\vskip1pc\noindent
What is different, however, is that the number of nodes in an eigenmode with given $m$ is no longer fixed. Consider, for instance, the unstable modes of the $3$-fold which come in a pair of opposite parity (see again Fig.~\ref{fig:unstablemodesneqthree}). The eigenmode has four nodes for $\se=\frac{\pi}{360}$ but eight for $\se=\pi$ and $2\pi$. A similar behavior is observed for the other $n$-folds (see Fig.~\ref{fig:unstablemodes} for the unstable modes of $n=4$). This makes it difficult, in general, to assign the correct $m$ to an eigenstate by inspection; its Fourier decomposition now displays a number of different frequencies 
(see Tab.~\ref{tab:eigenvectors} for the lowest modes of $n=2$ and $\se = 2\pi$).

\begin{table}
\begin{center}
\begin{tabular}{|c||c|p{13cm}|}
\hline && \\[-0.2cm]
m & eigenvalue & eigenmode \\[0.2cm]
\hline\hline&& \\[-0.4cm]
0  & 1.41 & $-0.2717 + 0.1075\cos{(4 s' )} - 0.0035\cos{(8 s' )} + 0.00008 \cos{(12 s' )}$\\
\hline && \\[-0.4cm]
$1^{-}$ & 0 &  $0.1581\sin{(s' )} + 0.3658\sin{(3 s' )} -0.0127\sin{(5 s' )}-0.0138\sin{(7 s' )}+ 0.0004\sin{(9 s' )}+0.0004\sin{(11 s' )} $\\
\hline && \\[-0.4cm]
$1^{+}$  & 0 & $0.1581\cos{(s' )} - 0.3658\cos{(3 s' )} -0.0127\cos{(5 s' )}-0.0138\cos{(7 s' )} + 0.0004\cos{(9 s' )}-0.0004\cos{(11 s' )} $\\
\hline && \\[-0.4cm]
2  & 0 & $-0.3984\cos{(2 s' )} + 0.0199\cos{(6 s' )} - 0.0005 \cos{(10 s' )}$ \\
\hline && \\[-0.4cm]
$3^{-}$ & 1.16&  $0.3649\sin{(s' )} - 0.1585\sin{(3 s' )} - 0.0292\sin{(5 s' )} + 0.0058\sin{(7 s' )}+ 0.0009\sin{(9 s' )}+0.0001\sin{(11 s' )} $\\
\hline && \\[-0.4cm]
$3^{+}$ & 1.16 &  $0.3649\cos{(s' )} + 0.1585\cos{(3 s' )} - 0.0292\cos{(5 s' )} - 0.0058\cos{(7 s' )}+ 0.0009\cos{(9 s' )}-0.0001\cos{(11 s' )} $\\
\hline && \\[-0.4cm]
$4^{-}$ & 6.08 & $0.3988\sin{(4 s' )} - 0.0120\sin{(8 s' )} + 0.0003 \sin{(12 s' )}$\\
\hline && \\[-0.4cm]
$4^{+}$ & 6.38 &  $-0.0760 - 0.3840 \cos{(4 s' )} + 0.0115 \cos{(8 s' )} - 0.0003 \cos{(12 s' )}$ \\
\hline && \\[-0.4cm]
$5^{-}$  & 22.74 &  $0.0319 \sin{(s' )} + 0.00003 \sin{(3 s' )} + 0.3975 \sin{(5 s' )} -
0.00002\sin{(7 s' )} - 0.0100\sin{(9 s' )}-0.0002\sin{(11 s' )} $\\
\hline && \\[-0.4cm]
$5^{+}$  & 22.74 &   $-0.0319 \cos{(s' )} + 0.00003 \cos{(3 s' )} - 0.3975 \cos{(5 s' )} -
0.00002\cos{(7 s' )} + 0.0100\cos{(9 s' )}-0.0002\cos{(11 s' )} $\\
\hline && \\[-0.4cm]
$6^{-}$ & 56.67 & $-44.4363 \Jacobisnadjustsquareb{1.8709 s'}{-0.3003} +
43.7022 \sin{(2 s')} -0.3299 \sin{(6 s')} +0.0033 \sin{(10 s')}$\\
\hline && \\[-0.4cm]
$6^{+}$ & 56.73 & $0.0199 \cos{(2 s' )} + 0.3984 \cos{(6 s' )} - 0.0086 \cos{(10 s' )}$\\
\hline
\end{tabular}
\end{center}
\caption{Eigenmodes and -values corresponding to the modes $m$ of the flat limit for $n=2$ and $\se=2\pi$. 
In the table the scaled variable $s' := \frac{2\pi s}{\st} = \frac{s}{2}$ is used.}
\label{tab:eigenvectors}
\end{table}
\vskip1pc\noindent
The natural resolution of this problem is to order the states with respect to increasing
eigenvalues. This provides a one-to-one correspondence which was also used to assemble
Tab.~\ref{tab:eigenvalues}. The zero modes, for example, are easily identified as the eigenvectors
that go together with the vanishing eigenvalues. Our numerical calculations are consistent with
the results of section~\ref{subsec:zeromodes}: one obtains the Fourier decomposition of three linearly 
independent eigenvectors each of which can be written as a linear combination of the three zero modes 
$\Phi_{\VECi}$,$\Phi_{\VECj}$, and $\Phi_{\VECk}$.
A fourth zero mode  analogous to the one found in the small surplus limit does not exist. However,
if isometry is relaxed in the numerical calculations, one finds a corresponding  eigenmode which
becomes unstable outside the small surplus limit. The Fourier series of this eigenmode displays
exactly the same frequencies as the Fourier  series~(\ref{eq:kappaFourierseries}) of the curvature
$\kappa$, $i. e.$, $m=(2l+1) n$  with $l\in \{0,1,2,\ldots\}$. If the coefficient of the lowest
frequency $m=n$ is set to the value of the corresponding coefficient of $\kappa$, all other
coefficients of the series of the mode are numerically found to be given by $2l+1$ times the
corresponding coefficient of the series of $\kappa$.

\vskip1pc\noindent The eigenmode $\Phi_\text{0}$ which corresponds to the constant mode ($m=0$) 
in the small surplus limit now gets distorted by the Fourier mode
$m=2n$ and small contributions of the multiples ($4n, 6n, \ldots$). $\Phi_\text{0}$ can be
identified by looking at its eigenvalue: for $\se\to 0$ the eigenvalue of $\Phi_\text{0}$ is given
by $n^2$  but diminishes if the surplus angle is increased (see again Tab.~\ref{tab:eigenvalues}).
For the surplus angles under consideration the eigenvalue is always positive, the state ($m=0$) is
thus stable.

\vskip1pc\noindent
One final observation concerns the eigenvalues: in the non-linear theory the two-fold degeneracy of
certain pairs gets lifted. From Tab.~\ref{tab:eigenvalues} one identifies these pairs as the stable
modes $m=4$ and 6 for the $2$-fold, $m=6$ for the $3$-fold, and for higher values of $m$ for the
$4$-fold. A quick glance in Tab.~\ref{tab:eigenvectors} provides an explanation in two distinct
sources.  One of these is the role of the constant term: whenever $m$ is a multiple of $2n$ the
eigenstate with even parity contains the constant term in its Fourier decomposition whereas the
corresponding eigenstate with odd parity does not. The second is the isometry constraint: the
Gram-Schmidt process breaks the symmetry between states of odd and even parity for those eigenstates
with $m=(2l+1)n$. All remaining eigenstates come in pairs with equal eigenvalues, exactly as in the
small surplus limit.


\section{\sf Conclusions}

We have examined the stability of unstretchable sheets which are free to bend. The conical defect
was studied in detail. Whereas cones with a surplus angle exhibit an infinite number of equilibrium
states, all but the ground state with $n=2$ are unstable. The dominant mode of instability will generally not
be directly towards the ground state; instead there will be a cascade of instabilities. 

\vskip1pc\noindent The fourth order differential operator, which arises
in our analysis of stability, also possesses mathematical properties that are of interest in their own
right. We have studied these properties to the extent that they touch on the issue of stability.

\vskip1pc\noindent Suppose that one removes a circular disk around the conical singularity at the center.
The conical annulus obtained possesses additional isometric modes of deformation. These deformations can be
decomposed into a conical part preserving the apex, as well as a part which destroys it. The resulting
surface will be a more general tangent developable. Interestingly, the linear operator determining the energy
of deformations is not diagonal with respect to this decomposition. It is no longer even physically obvious
that the two-fold conical ground state will continue to be stable with respect to isometric deformations
destroying the cone.

\vskip1pc\noindent  While we have limited our analysis to the behavior of an isolated conical defect, the
framework developed here has more general validity. Analogous operators will occur in the study of the stability
of related geometries: For example, for a d-cone subject to external constraints \cite{cerdamaha} the fourth 
order operator will have a form identical to Eqn.~(\ref{eq:operatorL}), with the potentials replaced by the relevant curvature 
function.
Discontinuities at the points of contact, however, will make the analysis more delicate. The extension of our
framework to sheets with a non-Euclidean metric \cite{Sharon_review,Santangelo} is also possible but will be less
straightforward: the analog of the operator ${\cal L}$ will be a partial differential operator; it will generally not be possible
to integrate out the radial dependence. In these geometries one can exploit the rotational symmetry which predicts
zero modes. When two or more defects are placed next to each other, however, one loses this symmetry and one would not
expect these modes to show up. This makes the problem more challenging from a technical point of view but does not present an 
obstacle in principle. In any case, our analysis in this paper is a reliable point of departure for further research. 

\vskip1pc\noindent  We have benefitted from discussions with
Martine Ben Amar, Osman Kahraman, and Norbert Stoop.
We thank the Aspen Center for Physics as well as the Kavli Institute for Theoretical Physics for their hospitality. Partial support from 
DGAPA PAPIIT grant IN114510-3 is acknowledged.

\begin{appendix}

\section{\sf  First order deformations $\delta K_{ab}$ and $\delta K$ for an isometric sheet} \label{app:1stvariationKKab}

In general, the first order deformation
of the extrinsic curvature is given by the covariant Hessian of the
deformation vector projected onto the normal:
\begin{equation}
\label{eq:delKab}
\delta K_{ab}= -\VECn\cdot \nabla_a \nabla_b \,\delta \VECX\,.
\end{equation}
In an isometry, the trace  $K$ deforms by a divergence:
\begin{equation}
\delta K = -\VECn\cdot \nabla^2 \delta \VECX = - \nabla_a ( \VECn\cdot \nabla^a \delta \VECX)\,.
\end{equation}
The first identity uses Eq.(\ref{eq:delKab}) with $\delta g^{ab}=0$. The second involves pulling $\VECn$ inside the covariant derivative;
the additional term generated, $K^{ab} \VECe_a\cdot \nabla_b \delta \VECX$,
vanishes for an isometry on account of the identity (\ref{delg0}).

\vskip1pc\noindent It is convenient to decompose $\delta \VECX$ into tangential and normal components:
\begin{equation}
\delta{\VECX}= \psi^a\,\VECe_a + \phi\,\VECn\,.
\end{equation}
Isometry (\ref{delg0}) places constraints on these components
\begin{equation}
\label{isomgen}
\nabla_a \psi_b + \nabla_b \psi_a + 2 K_{ab} \phi =0\,.
\end{equation}
\vskip1pc\noindent A simple calculation which makes use of the Gauss structure equations,
$\nabla_a \VECe_b= -K_{ab} \VECn$,
then gives
\begin{equation}
  \delta K =
  - \nabla^2 \phi + \nabla_a (K^{ab}\psi_b)
  \,.
\end{equation}
A useful alternative, non-divergence, expression for $\delta K$
is given by
\begin{equation}
\label{delK2}
\delta K=
(-\nabla^2 + 2K_G - K^2 )\phi + (\nabla_a K)\, \psi^a \,.
\end{equation}
Its counterpart for $\delta K_{ab}$ can be expressed as
\begin{equation}
\delta K_{ab} = \delta_\perp K_{ab} + \delta_\| K_{ab}\,,
\end{equation}
where
\begin{eqnarray}
\delta_\perp K_{ab} &=&  - \nabla_a \nabla_b \phi + K_{ac} K_{b}^{c}\, \phi \, ,\nonumber\\
\delta_\| K_{ab} &=&  K_{ac} \nabla_{b}\psi^c + K_{bc} \nabla_{a}\psi^c
+ \psi^c \nabla_c K_{ab}\,.
\end{eqnarray}


\section{\sf Derivation of Eq.(\ref{eq:del2HVs})} \label{app:2ndvarcone}

Modulo two
integrations by parts on the first term involving $\delta \kappa''$ in Eq.(\ref{eq:del2H}) we have
\begin{eqnarray}
\delta^2 H &=& \oint ds \, \Big\{ \Phi \Phi'''' + \frac{1}{2}
[2 + 3
\kappa^2 + 2 (1- C_\|)] \Phi \Phi '' + \frac{1}{2} [3\kappa^2 + 2(1-C_\|)]
\Phi^2
\nonumber\\
&& -  (\kappa \Psi )' \Phi '' -\frac{1}{2} [ 3\kappa^2 + 2(1-C_\|)] (\kappa
\Psi)' \Phi \Big\}\,.\nonumber
\end{eqnarray}
We now use the local constraint (\ref{eq:arccon}) to eliminate
the $\Psi$ dependence appearing in the last two terms $(I, II)$ in
favor of $\Phi$.
Note that the Euler-Lagrange equation ${\cal E}_\perp=0$ implies
\begin{equation}
\kappa''' = - \frac{1}{2}
[ 3\kappa^2 + 2(1-C_\|)]\kappa'\,. \label{eq:k3p}
\end{equation}
Using Eqs.(\ref{eq:arccon}) and (\ref{eq:k3p}),
it becomes possible to re-express the term $II$ as
\begin{equation}II=  -  \frac{1}{2} [ 3\kappa^2 +
2(1-C_\|)] (\kappa \Psi)' \Phi = \kappa''' \Psi \Phi + \frac{1}{2} [3
\kappa^2 + 2(1-C_\|)]\kappa^2 \Phi^2 \,.
\end{equation}
Modulo a total derivative,
\begin{equation} \kappa''' \Psi \Phi =  \kappa\kappa''\Phi^2 - \kappa''
\Psi \Phi'\,.
\end{equation}
On the other hand
\begin{equation}I= -  (\kappa \Psi )' \Phi ''=  \kappa^2 \Phi\Phi'' - \kappa' \Psi
\Phi''\,.
\end{equation}
We thus obtain (modulo total derivatives and the Euler-Lagrange equation)
\begin{eqnarray} I+II &=&
\kappa^2 \Phi\Phi'' - \Psi (\kappa'\Phi')' + \left[\kappa'' +
\frac{3}{2}\kappa^3 + (1-C_\|) \kappa \right] \kappa \Phi^2 \nonumber\\
&=&  \kappa^2 \Phi\Phi''  + \left(\frac{1}{2} \kappa\kappa'' + \frac{1}{2} (\kappa')^2
+ \kappa^4 \right)\Phi^2 \,.
\end{eqnarray}
Summing terms, Eq.(\ref{eq:del2HVs}) follows.

\end{appendix}

\end{document}